\documentclass[aps, prl, twocolumn,superscriptaddress]{revtex4-1}

\usepackage[utf8]{inputenc}
\usepackage{lipsum}
  
\usepackage{amsmath}
\usepackage{color}
\usepackage{graphicx}
\usepackage{epstopdf}
\usepackage{color}
\usepackage{pdfpages}
\usepackage{pgffor}

\makeatletter
\AtBeginDocument{\let\LS@rot\@undefined}
\makeatother

\begin{document}

\title{Simple computer program to calculate arbitrary tightly focused (propagating and evanescent) vector light fields}

\author{Isael Herrera}
\altaffiliation[Current address: Fresnel Institute, isael.herrera@fresnel.fr]{}
\author{Pedro A. Quinto-Su}%
\email{pedro.quinto@nucleares.unam.mx}
\affiliation{ Instituto de Ciencias Nucleares, Universidad Nacional Aut\'onoma de M\'exico, Apartado Postal 70-543, 04510, Cd. Mx., M\'exico.}

\begin{abstract}
In this work we present a simple code to calculate tightly focused vectorial light fields (propagating and evanescent) generated by input fields that have arbitrary amplitude, phase and polarization.
The program considers results from previous studies, like integration via fast Fourier transforms to speed up the integration. The calculations are done in a Cartesian coordinate system that is convenient to compare with experimental results for beams that are shaped with programmable optical elements like spatial light modulators or digital micromirror arrays. 
We also discuss how to avoid diverging terms at the origin by shifting the angular mesh by half a point and correcting the output by cancelling the phase term that arises from the shifted Fourier transform. 
\end{abstract}

\maketitle

\section{Introduction}
In general, tightly focused light has significant polarization components in the 3 spatial directions and can be described by the vectorial model that Richards and Wolf developed in 1956 \cite{rw}. 
Their model has become even more relevant, as some of the most important optical applications rely on tightly focused laser light. 

It is now possible to experimentally generate beams with arbitrary amplitude, phase and polarization, which upon tight focusing can result in very complex structures with rapid changes in amplitude and phase at subwavelength spatial scales in each polarization component. So, the ability to compare the measurements with the calculations while performing the experiment can help identifying the focal plane and possible errors in the input field (like the size at the aperture).
For example, for beams that have radial or azimuthal symmetry, simply overfilling the back aperture of the microscope objective might be sufficient to get good agreement with the calculation. However, when the input beams do not posses those symmetries, it is crucial to check the precise spatial dimensions and the equivalence between spatial and angular coordinates over the lens aperture. 

In general, the numerical integration in the Richards-Wolf model requires several minutes in a laptop computer to calculate the field at a single transverse spatial plane in the vicinity of the focus. A crucial step to speed up the calculation (by a factor $\sim 30-100$) has been the use of the fast Fourier transform algorithm to perform the numerical integration \cite{fastfocal}. 

Recently, a couple of open source codes that calculate tightly focused beams have been published: Infocus \cite{infocus} and Pyfocus \cite{pyfocus}, which have helped increase the availability of these tools to experimental groups. Infocus considers propagating fields and the calculations were compared with intensity measurements of beams focused with lenses with numerical apertures in the range between $0.4-0.7$. Pyfocus is more oriented toward microscopy and intensity distributions, it also considers straight boundaries and evanescent fields. However, both codes describe the fields in cylindrical coordinates and do not consider completely arbitrary vector fields at the input.
Also, there are still some details that can be improved in order to simplify the way arbitrary fields are described, matching the geometry of the experimental devices that are used to generate the beams and enable direct point by point comparisons with the experimental measurements. 

In this work we provide a simple MATLAB code based on previous results \cite{rw, fastfocal, fastfocalcartesian} to calculate arbitrary tightly focused vectorial fields. 
In the Code (Appendix) we replaced the traditional spherical coordinate system by Cartesian coordinates (spatial and angular) as in \cite{fastfocalcartesian, novotnybook}. This approach helps matching the description of arbitrary fields to those in the experiments, where the structured light is generated with computer controlled rectangular arrays like spatial light modulators (SLM) or digital micromirror arrays (DMD). Furthermore, Cartesian coordinates are the most suitable to express the beams as a superposition of plane waves (in the angular spectrum representation) and for the Fast Fourier transform.

The article is presented in the following way: we start with a summary of the Richards-Wolf model in section 2, then we present a simplified optical system to generate these beams and how it relates to the calculation. In section 4 we describe the numerical implementation which is based on \cite{fastfocal} and discuss in detail the construction of the integration mesh, the correction of artifacts and the spatial resolution associated with the physical and computational parameters. This discussion is also useful for simulating diffraction and focusing of paraxial light fields. Section 5 contains several examples with different input polarization states including the small difference between a corrected and uncorrected field which is a mistake which is commonly made. Finally, in section 6 we present the case of a planar interface with an example. The computer programs are in the Appendices.

\section{General Theory: Richards Wolf model}
The model of Richards and Wolf \cite{rw} is derived from energy conservation, the sine condition, and the fact that the field at the lens surface (represented by conjugate rays in geometrical optics) corresponds to the far field $\textbf{E}_\infty$ of the focused field $\textbf{E}$. 

The incident beam ($\textbf{E}_{inc}$) is focused by an aplanatic lens (Figure 1).
Inside the spherical surface that focuses the light, the refractive index is $n_2$ ($n_2=1.518$ for glass and immersion oil). We assume the incident light propagates from left to right in air ($n_1 =1$) with a transverse polarization state before it is focused by the lens. Here we only consider the light that is transmitted by the lens which is focused, so we assume an unitary transmission coefficient.
Upon focusing, each transverse polarization component acquires additional components on the other spatial directions. 
The fields are described using spherical coordinates which naturally match the directions of the parallel and perpendicular components of the electric fields to the surface of the aplanatic lens.

 \begin{figure}[t]
    \centering
    \includegraphics[width=.40\textwidth]{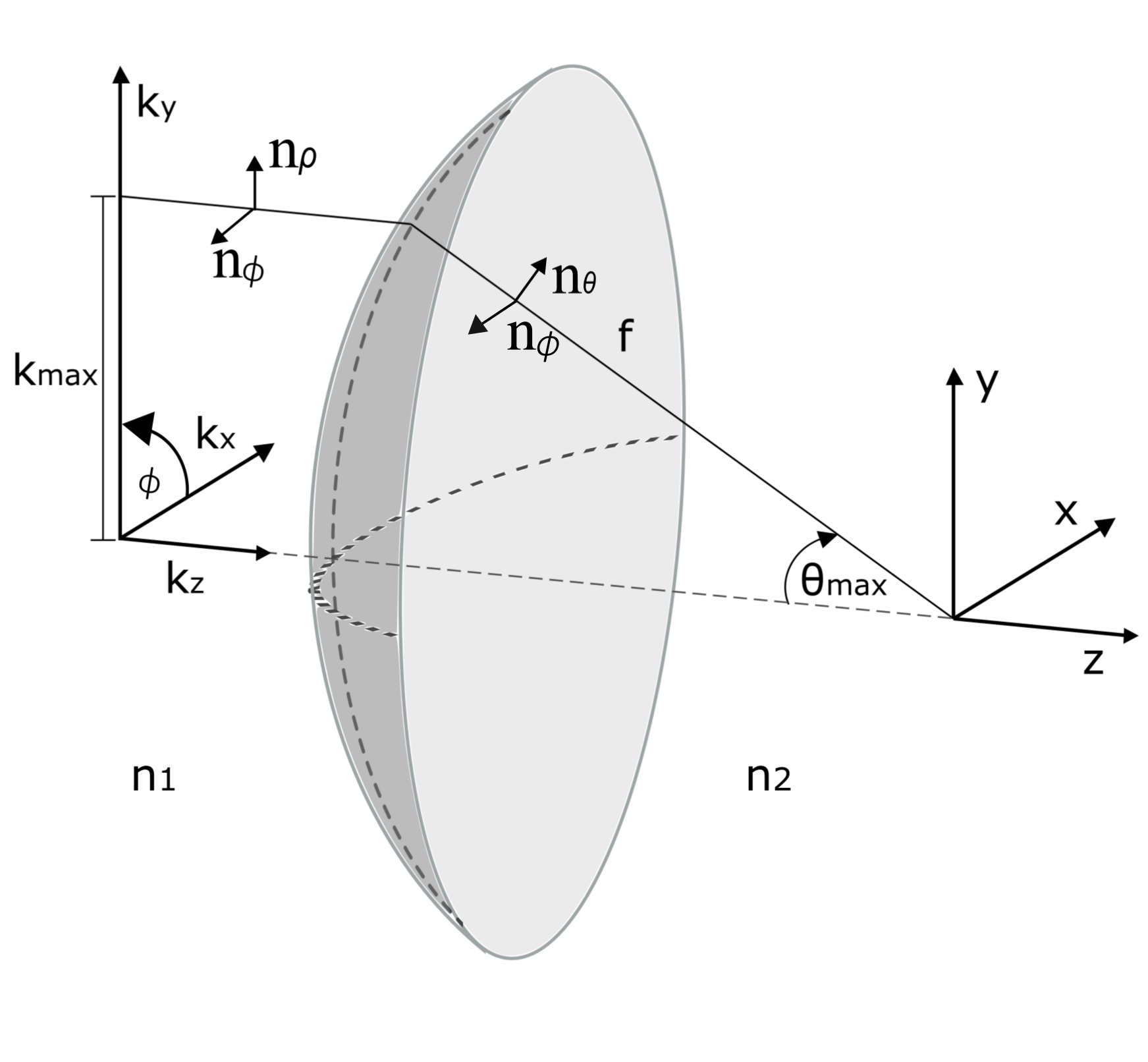}
    \caption{Representation of the focusing model. A paraxial beam propagating in a medium with refractive index $n_1$, incident into lens that focuses at $f$ in a medium with a refractive index $n_2$. The lens is represented by a spherical surface of reference centered at the focal point with radius $f$. }
\end{figure}


The relations between the spherical and Cartesian unitary vectors are:

\begin{equation}
\begin{split}
\textbf{n}_{\phi}&=-\sin(\phi)\textbf{n}_x+\cos(\phi)\textbf{n}_y\\
\textbf{n}_{\rho}&=\cos(\phi)\textbf{n}_x+\sin(\phi)\textbf{n}_y\\
\textbf{n}_{\theta}&=\cos(\theta)\cos(\phi)\textbf{n}_x+\cos(\theta)\sin(\phi)\textbf{n}_y-\sin(\theta)\textbf{n}_z
\end{split}
\end{equation}

If the incident field is linearly polarized in the $x$ direction, it can be expressed as  $E_{inc}^x=E_{0x} e^{i \phi _{0x}} \textbf{n}_x$, where $E_{inc}^x$ and $\phi _{0x}$ are the amplitude and the phase and $\textbf{n}_x$ is the unitary vector in the $x$ direction.
Upon focusing, the resulting field has polarization components along the 3 spatial directions. The far field representation of the x component of the focused field as a function of the incident field is $E_\infty ^x $ which has the 3 polarization components (in a Cartesian basis).

\begin{align}
  \textbf{E}_{\infty}^{x}= \sqrt{\frac{n_1}{n_2}}\sqrt{\cos(\theta)}E_{inc}^x  \Bigg\{  -\sin(\phi)  \begin{pmatrix}
           -\sin(\phi) \\
            \cos(\phi)\\
           0
         \end{pmatrix}  + \notag\\
         \cos(\phi)  \begin{pmatrix}
           \cos(\theta)\cos(\phi) \\
            \cos(\theta)\sin(\phi)\\
           -\sin(\theta)
         \end{pmatrix}   \Bigg\}  
\end{align}

While for the case of an incident field $\textbf{E}_{inc}=E_{inc}^{y}\textbf{n}_y$ with $E_{inc}^y=E_{0y} e^{i \phi _{0y}}$. The $E_\infty ^y $ representation of the focused field is:
\begin{equation}
\begin{split}
  \textbf{E}_{\infty}^{y}=\sqrt{\frac{n_1}{n_2}}\sqrt{\cos(\theta)}E_{inc}^y  &  \Bigg\{ cos(\phi)  \begin{pmatrix}
-\sin(\phi) \\
 \cos(\phi)\\
0
\end{pmatrix}  + \\ & \sin(\phi)  \begin{pmatrix}
\cos(\theta)\cos(\phi) \\
\cos(\theta)\sin(\phi)\\
-\sin(\theta)
\end{pmatrix} \Bigg\} 
\end{split} 
\end{equation}

Equations 2 and 3 can be rewritten in Cartesian coordinates using:
\begin{equation}
\begin{split}
\frac{x_{\infty}}{f}&=\frac{k_x}{k}  =\sin(\theta)\cos(\phi)
\\
\frac{y_{\infty}}{f}&=\frac{k_y}{k} =\sin(\theta)\sin(\phi)
\\
\frac{z_{\infty}}{f}&=\frac{k_z}{k}  =\cos(\theta)
\end{split}
\end{equation}
where $k=n k_0$ ($k=k_1$ in sections 4-6) and $k_0 =2\pi /\lambda $, where $n$ is the refractive index and $\lambda$ the wavelength in vacuum.

The expressions for $\textbf{E}_{\infty}^x$ and $\textbf{E}_{\infty}^y$ are:
\begin{equation}
\textbf{E}_{\infty}^x=\sqrt{\frac{n_1}{n_2}}\sqrt{\frac{k_z}{k}} E_{inc}^x(\frac{k_x}{k},\frac{k_y}{k})  \begin{pmatrix}
           k_y^2+k_x^2 k_z/k\\
            -k_x k_y +k_x k_y k_z /k\\
           -(k_x^2+k_y^2)k_x/k)
         \end{pmatrix} \left( \frac{1}{k_x^2+k_y^2} \right)
         \label{eqn:lentefactorx}
\end{equation}

For the other transverse polarization component (direction $\textbf{n}_y$). \\
\begin{equation}
\textbf{E}_{\infty}^{y}=\sqrt{\frac{n_1}{n_2}}\sqrt{\frac{k_z}{k}} E_{inc}^y(\frac{k_x}{k},\frac{k_y}{k})  \begin{pmatrix}
           -k_x k_y+k_x k_y k_z/k\\
            k_x^2+k_y^2 k_z /k\\
           -(k_x^2+k_y^2)k_y/k)
         \end{pmatrix} \left( \frac{1}{k_x^2+k_y^2} \right)
         \label{eqn:lentefactory}
\end{equation}

Finally the focused field at $(x,y,z)$ can be written as:
\begin{equation}
\begin{split}
\textbf{E}(x,y,z)=-\frac{ik f e^{-i k f}}{2\pi}\iint \limits_{k_x^2+k_y^2\leq k_{max}^2} \left[ c_x \textbf{E}_{\infty}^{x}+c_y\textbf{E}_{\infty}^{y} \right]\frac{1}{k_z} \times \\
e^{i\left[k_x x+k_y y+k_z z \right]}d k_x d k_y
\end{split}
\label{eqn:resfin}
\end{equation}
where $z=0$ represents the focal plane. In general $c_x$ y $c_y$ are complex 2d functions that depend on  $(k_x,k_y)$ and define the polarization state of the light at the back aperture of the lens. In the cases of linear or elliptical polarization states, $c_x$ and $c_y$ are constants over the aperture domain. 
The integration domain is restricted to spatial frequencies that satisfy $k_x^2+k_y^2\leq k_{max}^2$, where in terms of the numerical aperture ($NA=n\sin \theta _{max}$) as $k_{max}=NA k_0$. The same limit of $k_{max}$ can also be expressed in terms of the aperture radius $R$ using equation 4: $R=k_{max} f/k=NA f/n$ (Fig. 1).

In order to extend the domain of the integrals in eq. (7) to infinity, an aperture function is needed that vanishes for frequencies larger than the magnitude of k:

\begin{equation}
\Theta(k_x,k_y)=
\left\{
	\begin{array}{ll}
		1 & \mbox{if } k_x^2+k_y^2 \leq k_{max}^2 \\
		0 & \mbox{if }  k_x^2+k_y^2 > k_{max}^2 
	\end{array}
\right.
\end{equation}

\begin{equation}
\begin{split}
\textbf{E}(x,y,z)=-\frac{ife^{-ikf}}{2\pi}\iint\limits_{-\infty}^{\infty}\Theta(k_x,k_y)\left[ c_x \textbf{E}_{\infty}^{x}+c_y\textbf{E}_{\infty}^{y} \right]\frac{1}{k_z} \times \\
e^{i\left[k_x x+k_y +k_z z \right]} dk_x dk_y
\end{split}
\end{equation}

The previous expression can be rewritten as:
\begin{equation}
\textbf{E}(x,y,z)=-\frac{ife^{-ikf}}{2\pi} \textsf{IFT} \left[ \Theta(k_x,k_y)\left[ c_x \textbf{E}_{\infty}^{x}+c_y\textbf{E}_{\infty}^{y} \right]\frac{1}{k_z} e^{ik_z z}  \right] 
\label{eqn:fftlens}
\end{equation}
which can be calculated with the Fast Fourier Transform algorithm (fft) as shown in \cite{fastfocal}.

\section{Optical system}
The simplified system that we consider is very similar to most beam shaping experimental setups (i.e. holographic optical tweezers) that use beam shaping elements like SLMs or DMDs that can modulate phase and/or amplitude. 
A vectorial beam input field with an arbitrary polarization state in the transverse direction ($xy$) can be prepared splitting the components, modulating them independently and then recombining using an interferometer setup where both components have the same optical path length (this can be done with one or two beam shaping elements). Then, a complex polarization mask can be added to set the polarization, this can be done at the beam shaping element or with polarization elements like a quarter wave plate or a q plate.
In general, the vector beam is focused by a lens that is placed at a distance of the focal length from the beam shaping element (located in the Fourier plane of the lens), then a second lens collimates the beam, projecting the surface of the beam shaping element to the back aperture of the microscope objective where the field $E_{inc}$ is described. This sets the spatial dimension of the incident field, so it also can be written as a function of $(x_\infty , y_\infty)$ (which is a resized projection of the beam shaping element), in addition to the descriptions in terms of the dimensionless $(k_x/k, k_y/k)$ or $(x_\infty /f,\, y_\infty /f)$. This means that at each point $(x_\infty , y_\infty)$, we know the amplitude, phase and polarization.

In the computer code, we follow the steps of the experiment: The input beam is decomposed into the two complex transverse components $E_{inc, x} e^{i \phi _x}$ and $E_{inc, y} e^{i \phi _y}$ and we define them as a function of the spatial coordinates.
Then we add a 2d complex polarization mask $c_x ,\, c_y$ to each component to project the beam into arbitrary polarization states. Notice that $c_x ,\, c_y$ can also be included in the definition of $E_{inc}$ but we decided to separate them to mimic the experiment. Finally there is the aperture $\Theta$ that can also be controlled by the beam shaping elements and limited by the lens.
Those 7 transverse 2d masks ($E_{incx}$, $E_{incy}$, $\phi _{incx}$, $\phi _{incy}$, $c_x$, $c_y$, $\Theta$) are drawn in Fig. 2 and represent $E_{inc}$ at the back aperture of the microscope objective. 


 \begin{figure}[t]
    \centering
    \includegraphics[width=0.40\textwidth]{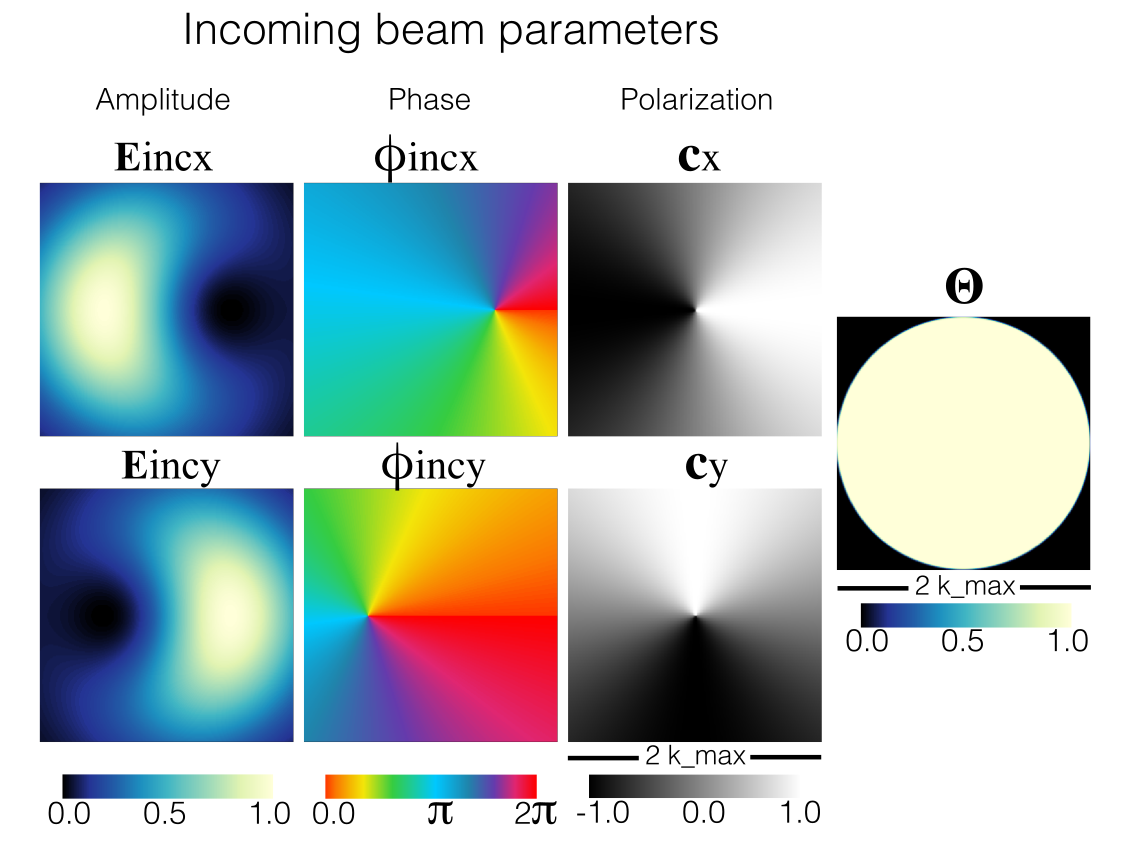}  
    \caption{Input field. The input plane where the incident field is described, amplitude, phase, polarization and aperture can be defined as functions of the angular representation $(k_x, k_y)$ or can also be written in terms of the spatial coordinates $(x_\infty , y_\infty)$. In this way, the points that span the aperture have a radius $R$ or $k_{max}$. 
    }
\end{figure}

In most experiments, one polarization component is modulated by a spatial light modulator than can control phase and amplitude, that modulated beam then propagates through a polarization element that controls $c_x$ and $c_y$ like a half wave plate (HWP) to control the a linear polarization state, or a quarter wave plate (QWP) for elliptical polarization. Radial and azimuthal polarization can be obtained with a vortex phase plate. Arbitrary beams can be implemented by modulating independently both transverse polarizations.
Fig. 2 shows the case for $E_{x0}=E_{y0}$ and $\phi _{x0}=\phi_{y0}$ and a spatially dependent polarization.

\section{Numerical Implementation}
To calculate the tightly focused field we consider 3 coordinate systems: the spatial Cartesian coordinates before the lens $(x_{\infty},y_{\infty})$ that describe the size of the aperture and the beam, the angular system at the same plane $(k_x,k_y)$ and the Cartesian system at the focal plane $(x,y)$. All coordinate systems are implemented in a grid of $L\times L$ points, where $L$ is an even number (for the FFT algorithm, usually $L=2^{11}-2^{12}$) to reduce errors with the numerical implementation of the Fourier Transform \textsf{FFT}.

In the following subsections we define the most important parameters of the focusing system like the numerical aperture and the effective focal length of the lens. Then grid domain,  spatial coordinates at the back aperture $x_\infty , y _\infty$, followed by the aperture function, the angular spectrum and finally the integration through a fft and the spatial grid at the focus.

\subsection{Focusing lens}
The focusing lens is usually a high numerical aperture microscope objective (oil immersion) with an effective focal length of $f$ and a numerical aperture $NA$, where $NA=n_2 R/f$ with $R$ the radius of the back aperture and $n_2$ the refractive index of the media in front of the lens ($n_2=1.518$ for immersion oil, $n_1=1$ for air). 
The effective numerical aperture $NA$ (can be adjusted with the beam shaping elements) or with a mechanical iris. 
We consider that the  back aperture is centered at the simulation domain and that it has a diameter of $N$ points (with $N$ an even number). 
Also, the area surrounding the aperture (with diameter $D$) has to be padded with zeros. A complete discussion of the expected errors as a function of the number of points for $L$ and $N$ respectively is in \cite{fastfocal}.

\subsection{Mesh}
As mentioned in the previous subsection, the grid has a size of $L\times L$ points or pixels. The 2 dimensional Fourier Transform is defined in that grid. In Matlab, the zero frequency is at the pixel coordinate $(L/2+1,L/2+1)$. When defining the spatial and angular coordinate system at the back aperture (same $L\times L$ grid), we set the origin of the coordinates at $(L/2+m_0,L/2+m_0)$ with $m_0=0.5$ to eliminate the zero frequency contributions that diverge in equations 5 and 6 (terms $1/(k_x^2 + k_y^2)$). Hence, the spatial and angular domains are $(-(L-1)/2, (L+1)/2)\Delta x $ and $(-(L-1)/2, (L-1)/2 )\Delta k$, with $\Delta x$ and $\Delta k$ the spatial and angular step sizes respectively and are defined in the following subsections. As a result, the Fourier transform is equal to the centered Fourier transform times a phase term $\phi_{shift}(X,Y)=- i 2\pi \frac{m_0}{L} X+ i 2\pi \frac{m_0}{L} Y $. Where $X$ and $Y$ are the dimensionless Cartesian coordinates centered at $(L/2+1,L/2+1)$. In order to correctly compute the complex electric field we have to multiply the resulting field components by the term $-\phi_{shift}(X,Y)$.

Figure 3 shows a comparison between the uncorrected (top row) and the corrected phase (bottom row) in the dominant polarization component of a highly focused Gaussian beam that is linearly polarized before focusing. That phase is well known and it consists of concentric constant sections that have a phase difference of $\pi$. We observe that the sections of the uncorrected phase are not constant, but have a small gradient which is more noticeable in the cross section plot (right column).

  \begin{figure}[t]  
    \centering
    \includegraphics[width=.5\textwidth]{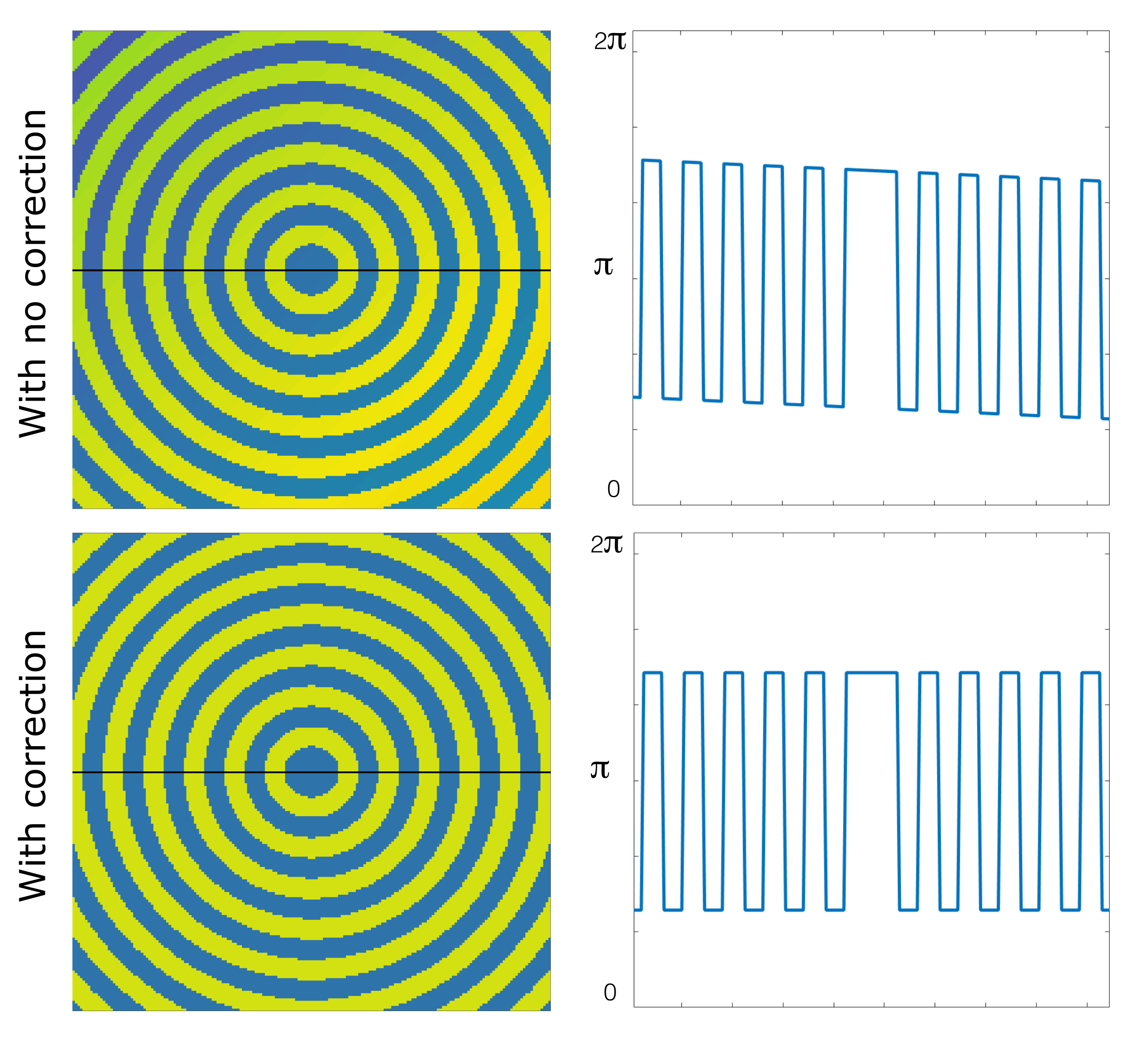}
    \caption{Effect of the origin translation in the phase of the resulting complex fields and correction. In the first column it is shown the phase of the x component of a focused x-linearly polarized Gaussian beam, in the second column it is shown the phase evaluated on the horizontal axis indicated on the phase map. }
    \label{fig:shift}
    \end{figure}  

\subsection{Spatial domain ($x_\infty, x_\infty$)}
At the back aperture of the microscope objective the spatial domain $x_\infty$ and $x_\infty$ is defined multiplying the discrete point grid by the constant $\Delta x_\infty$ which converts it into a spatial grid (dimensions of micrometers in the code). $\Delta x_\infty$ is defined by the size of the aperture which is defined with a radius of $N$ ($N=2^7$) points which is equivalent to the physical size of the aperture radius $R$ (in microns). In this way $\Delta x_\infty =R/N$. The spatial grids are $(-(L-1)/2, (L-1)/2) \Delta x_\infty$ ($m_0=0.5$). 

\subsection{Angular spectrum $(k_x, k_y)$}
The angular spectrum coordinates are defined as $k_x = k_1 x_\infty/f$ and $k_y = k_1 y_\infty/f$, where $k_1=n_2 k_0$ and $k_0 =2\pi/\lambda$. The size of each grid point in the angular spectrum mesh is $\Delta k=k_0 NA/N$. Notice that the interval does not include zero, this is done to avoid diverging terms $1/(k_x ^2+k_y ^2)$. 

After applying the fft2, the size of the spatial output mesh is defined by $\Delta x_f= 2\pi/(L \Delta k)=N \lambda/(L (NA))$ ($\Delta x_f =\Delta y_f$).

\subsection{Input Field $E_{inc}$ and polarization mask}
The input field with the polarization state is described by the seven 2-dimensional masks in Fig. 2: amplitudes $E_{x0}$, $E_{y0}$, phases $\phi _{x0}$, $\phi _{y0}$, aperture $\Theta $ and polarization $c_x$, $c_y$; all defined on the $L\times L$ domain. The polarization mask at a given point can be represented by $a e^{ib}$, with $\textsf{abs}(a)=1$.
As we mentioned before, in the experiments sometimes it is more convenient to use the spatial coordinates $(x_\infty , y_\infty)$ to define the input field, because the size of the field is defined at the beam shaping which is projected onto the back aperture of the microscope objective. Also the spatial dimensions of the input field can be measured with a CCD camera.

Hence, in the examples of section 5, we choose to describe the input as a function of the spatial coordinates.

\subsubsection{Aperture (circular, square, annular)}
The step function in eq. 8 depends on the angular spectrum. However, we can also express it in terms of its spatial size $R$ which is convenient because is the way it is measured.
When implementing an \textsf{FFT} it is very important to consider that a sudden discontinuity will result in the appearance of oscillations or speckle (aliasing). Those effects can be minimized by changing the definition of the aperture function. Here we present the result of \cite{fastfocal} it in terms of $(x_\infty, y_\infty)$, in the case of a circular aperture:

\begin{equation}
\Theta'(x_\infty,y_\infty)=\frac{1}{2}\left(1+\tanh\left[\frac{3}{4\Delta X_\infty}\left( R-(x_\infty^2+y_\infty^2)^{1/2}  \right) \right] \right)
\end{equation} 
where $R=N\Delta X_\infty =f (NA)/n_2$.

Other common geometries for the aperture are a square or a ring which we can describe in the other angular system to show the equivalence. 
In the case of a square aperture, the continuous version of the step function $\Theta (k_x,k_y)$ is: \begin{equation}
\begin{split}
\Theta (k_x,k_y)=\frac{1}{4}\left(1+\tanh\left[\frac{3}{4\Delta k}\left(\frac{k_{max}}{\sqrt{2}}-k_x  \right) \right] \right) \times\\ \left(1+\tanh\left[\frac{3}{4\Delta k}\left(\frac{k_{max}}{\sqrt{2}}-k_y  \right) \right] \right)
\end{split}
\end{equation}
where the $\sqrt{2}$ factor dividing $k_{max}$ appears because the maximum diameter for a square aperture is the diagonal. 

The modified $\Theta (k_x,k_y)$ for an annular aperture is:
\begin{equation}
\begin{split}
\Theta (k_x,k_y)=\frac{1}{2}\tanh\left[\frac{3}{4\Delta k}\left( k_{max1}-(k_x^2+k_y^2)^{1/2}  \right) \right] -\\
\frac{1}{2}\tanh\left[\frac{3}{4\Delta k}\left( k_{max2}-(k_x^2+k_y^2)^{1/2}  \right) \right]
\end{split}
\end{equation} 
where $k_{max}\geq k_{max1}>k_{max2}$ and $\Delta k_t =k_{max1}-k_{max2}$ is the width of the annular aperture.

\subsection{Integration}
Once the mesh, coordinates and input field set, then the argument of eq. (10) is written and the Fast Fourier Transform is calculated. 

     \begin{figure}[t]  
    \centering
    \includegraphics[width=.43\textwidth]{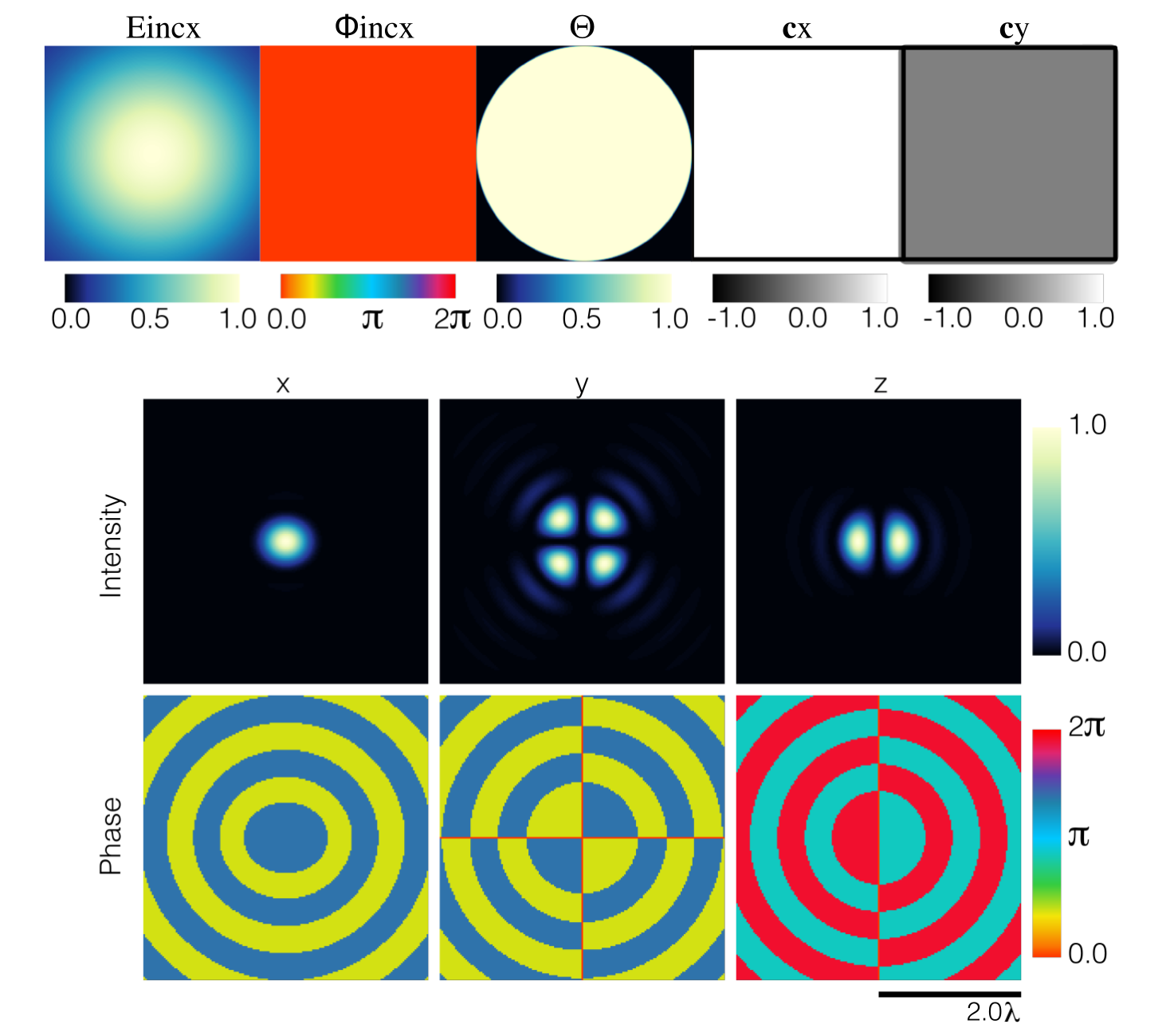}
    \caption{Focusing a linearly polarized Gaussian beam. The parameters are: $L=2^{12}$, $N=2^7$, $n_2=1.518$, NA$=1.3$, $f=2$ mm.    }
    \label{fig:simple}
    \end{figure}  

\section{Examples}
In the following subsections we show specific examples with different polarization states and input fields. The incident fields are described in the spatial or the angular coordinates. The magnitude of the errors and the dependency on the domain size are discussed in \cite{fastfocal}.

\subsection{Linearly polarized Gaussian}
We consider a Gaussian beam defined as: 
\begin{equation}
E_{inc}(x_{\infty},y_{\infty})=E_0 e^{\frac{-\left(x_{\infty}^2+y_{\infty}^2 \right)}{w_0^2}}e^{i\phi(x_{\infty},y_{\infty})}
\end{equation}
where $E_0$ (set to 1) is the field amplitude before the lens, $w_0$ is the waist of the Gaussian, $R=D/2$ is the aperture radius and $f_0$ the filling factor for a Gaussian beam. The filling factor is defined as $f_0=\omega_0 / R$, with  $\omega_0$ the Gaussian waist. 
The same field in angular coordinates:
\begin{equation}
E_{inc}(k_x,k_y)=E_0 e^{\frac{-(k_x^2+k_y^2)}{w'^2_0}}e^{i\phi (k_x,k_y)}
\end{equation}
where the new waist is $w'_0=w_0 k_1/f$.

Figure  \ref{fig:simple} shows the results for a Gaussian beam initially polarized in the direction $\textbf{n}_x$, focused by a lens with $NA=1.3$, $n=1.518$, $f=2$ mm, $f_0=1.0$ and $R=  1.71$ mm.  \\
In the case of a collimated Gaussian beam, it is a good approximation to consider the phase as constant, so it is set to $\phi(x_{\infty},y_{\infty})=\phi(k_x,k_y)$. 

Then, a linear polarization state is selected with the $c_x$ and $c_y$ coefficients. In this case we select only the $x$ component defining $c_x=1$ and $c_y=0$ in the $L\times L$ domain.

The top row in Fig. 4 has the 2 dimensional amplitude, phase, aperture, x polarization and y polarization masks. 
The 6 frames in the bottom depict the result at the focal plane with the 3 intensities and phases.

 \begin{figure}[t]  
    \centering
    \includegraphics[width=.425\textwidth]{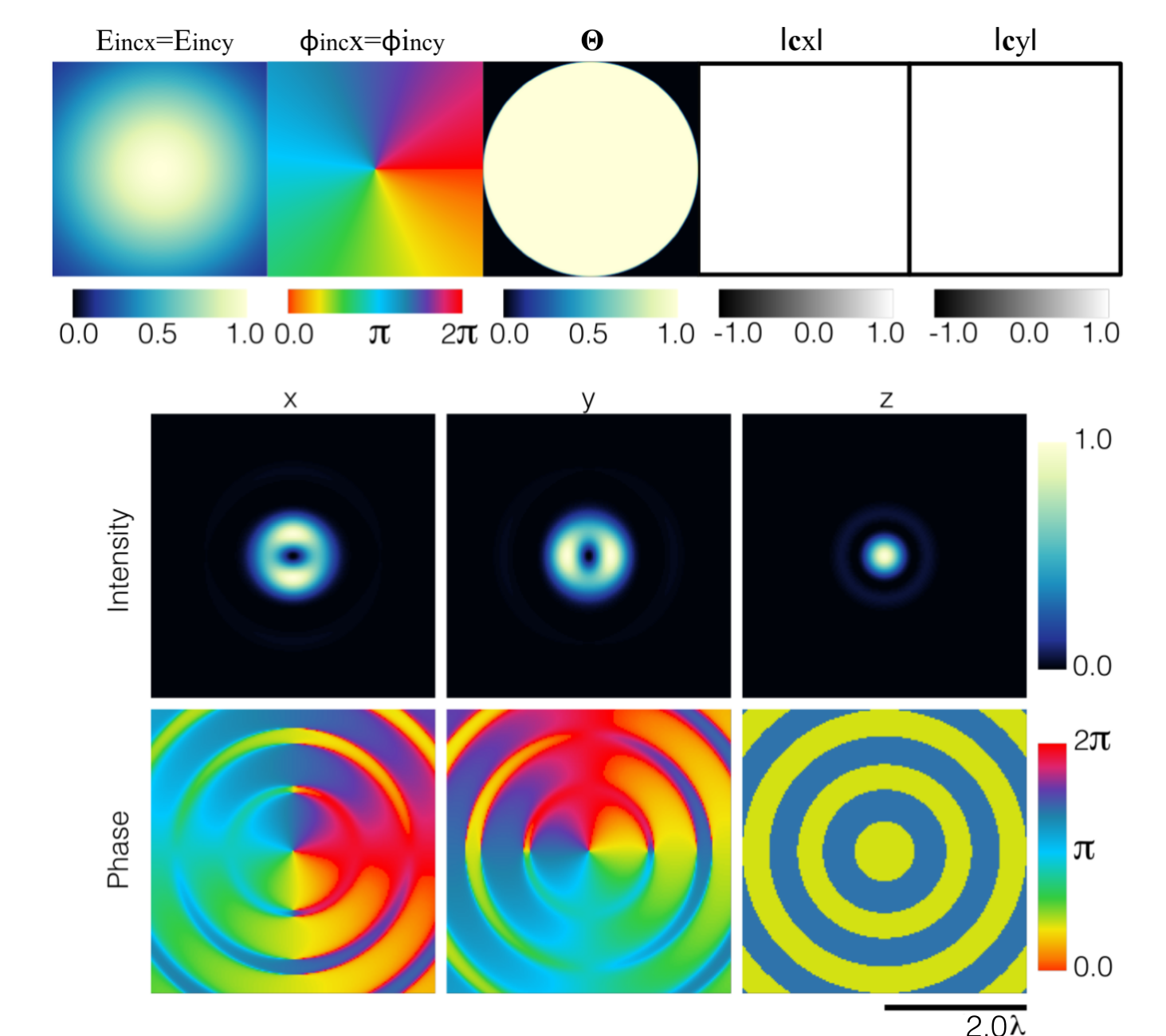}
    \caption{Focusing a Gaussian beam right circular polarization, $\sigma=1$, and vortex phase of order $m=-1$. The parameters are: $L=2^{12}$, $N=2^7$, $n_2=1.518$, NA$=1.3$, $f=2$ mm.   }

    \end{figure}  

\subsection{Circular polarization}
Same Gaussian with an optical vortex phase with $\phi_{0x}=\phi _\infty m$ $m=-1$ through a quarter wave plate, where $\phi _\infty$ is the azimuthal angle in the $x_\infty , y_\infty$ plane.
The polarization state, which is defined as $c_x =1$, $c_y=i$ (left circular polarization with $\sigma =1$) in the $L\times L$ domain.
The input planes in the domain $x_\infty$, $y_\infty$ (same plane as $k_x$, $k,y$) are at the top of Fig. 5, while the resulting field at the focus with the 3 intensities and phases is in the lower part. 
We observe that the in the $z$ component there is coupling between the orbital $m$ and the polarization $\sigma$ resulting in a null charge ($m=0$).

\subsection{Radial Polarization}
Figure 6 shows the polarization masks for the transverse polarization components with $c_x=\cos{\phi} = k_x/\sqrt{k_x^2+k_y^2}$ and $c_y=\sin{\phi} = k_y/\sqrt{k_x^2+k_y^2}$. 
We introduced auxiliary polar coordinates $\phi$ and $\rho$ in the $x_\infty$, $y_\infty$ domain. The angle $\phi$ can also be defined as the inverse tangent of $k_y/k_x$.
The calculation considers an annular mask and the same Gaussian input.

 \begin{figure}[t]  
    \centering
    \includegraphics[width=.4\textwidth]{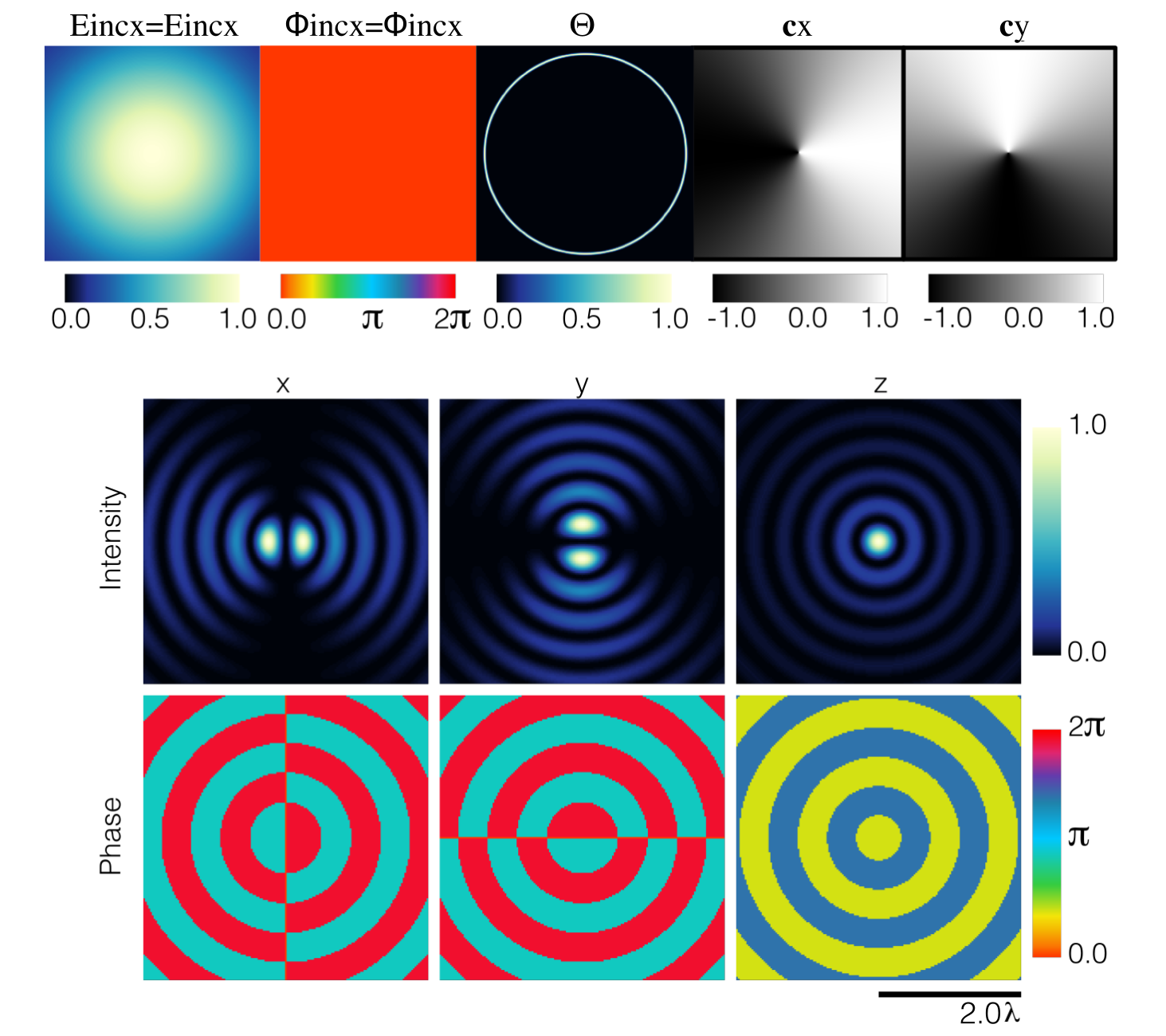}
    \caption{Focusing a radial polarized annular aperture (TM 0th).  The parameters are: $L=2^{12}$, $N=2^7$, $n_2=1.518$, NA$=1.3$, $f=2$ mm.   }

    \end{figure}  

\subsection{Azimuthal Polarization}
Figure 7 shows the polarization masks for the transverse polarization components $cx=\cos{\phi}=-k_y/\sqrt{k_x^2+k_y^2}$ $c_y=\sin{\phi}= k_x/\sqrt{k_x^2+k_y^2}$.
This case is interesting because there is no axial component as expected for this case, the ratio between the amplitude of the $E_z$ and the $E_x$ has an order of $10^{-19}$.

 \begin{figure}[t]  
    \centering
    \includegraphics[width=.4\textwidth]{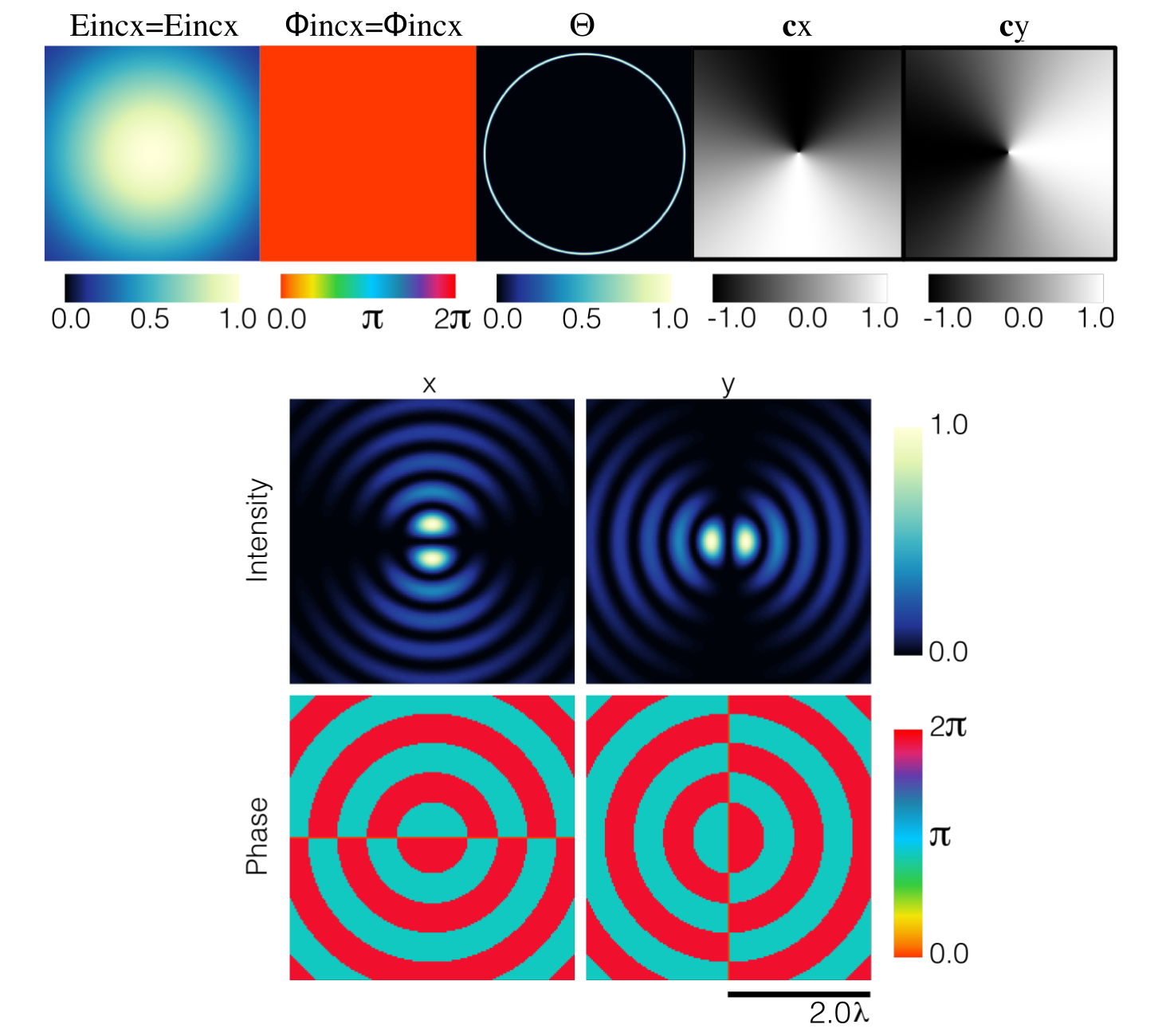}
    \caption{Focusing a azimuthal polarized annular aperture (TE 0th). The parameters are: $L=2^{12}$, $N=2^7$, $n_2=1.518$, NA$=1.3$, $f=2$ mm.   }

    \end{figure}  

\subsection{Arbitrary Polarization}
We also consider a Gaussian beam with flower (spider web) polarization state as described in \cite{denz} defined by the parameter $s=8$ $(-8)$:
The polarization is $c_x=\cos{((s/2)\phi))}$ and $c_y=\sin{((s/2)\phi))}$ with $s=8$, (flower). This is shown in Fig. 8. 
Inverting the sign in $s$ yields a similar pattern (spiderweb).

 \begin{figure}[t] 
    \centering
    \includegraphics[width=.4\textwidth]{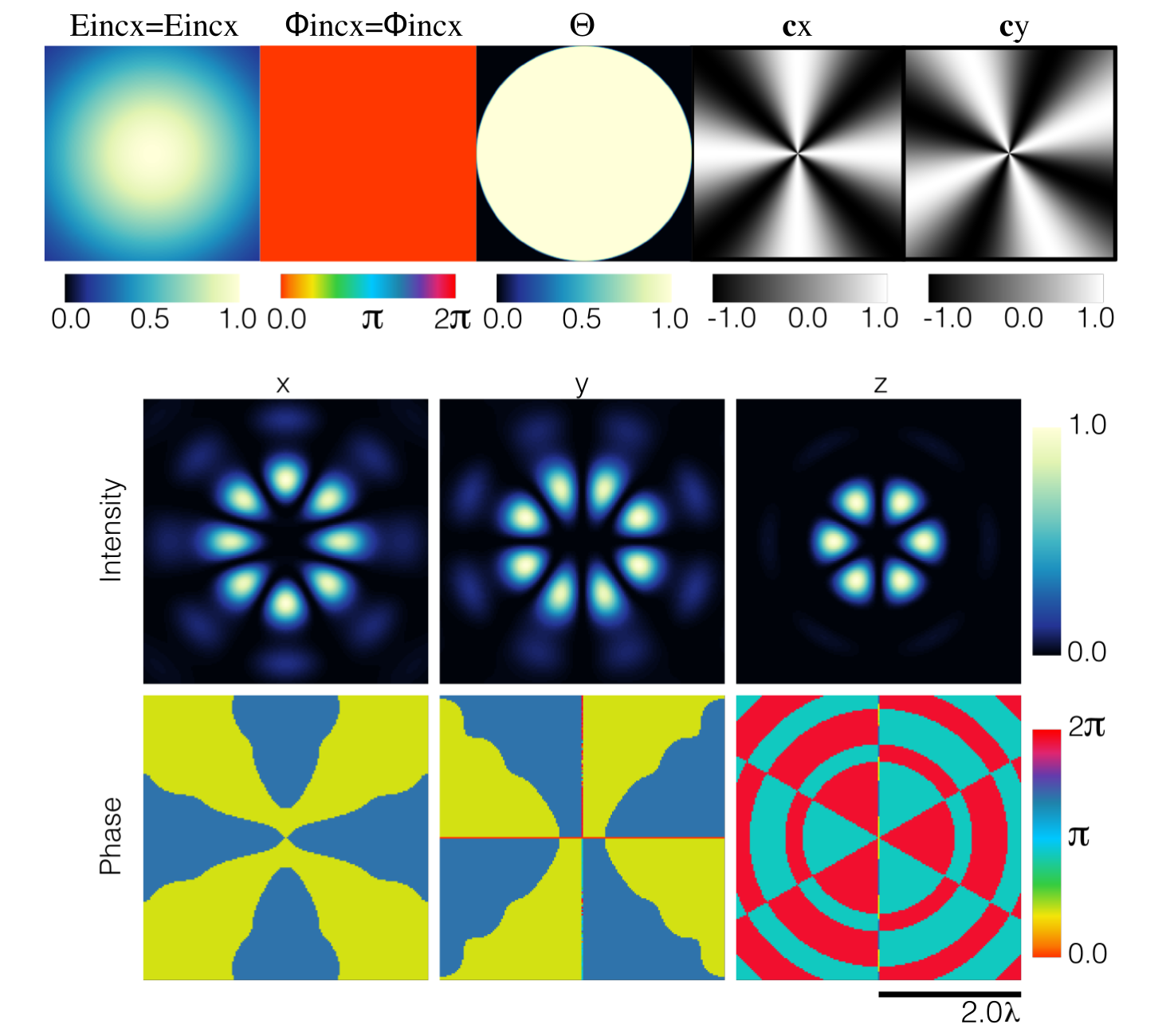}
    \caption{Focusing a Gaussian beam with flower (order 8) polarization: The parameters are $L=2^{12}$, $N=2^7$, $n_2=1.518$, NA$=1.3$, $f=2$ mm.   }

    \end{figure}  

\subsection{Arbitrary beam}
\begin{figure}[h] 
    \centering
    \includegraphics[width=.4\textwidth]{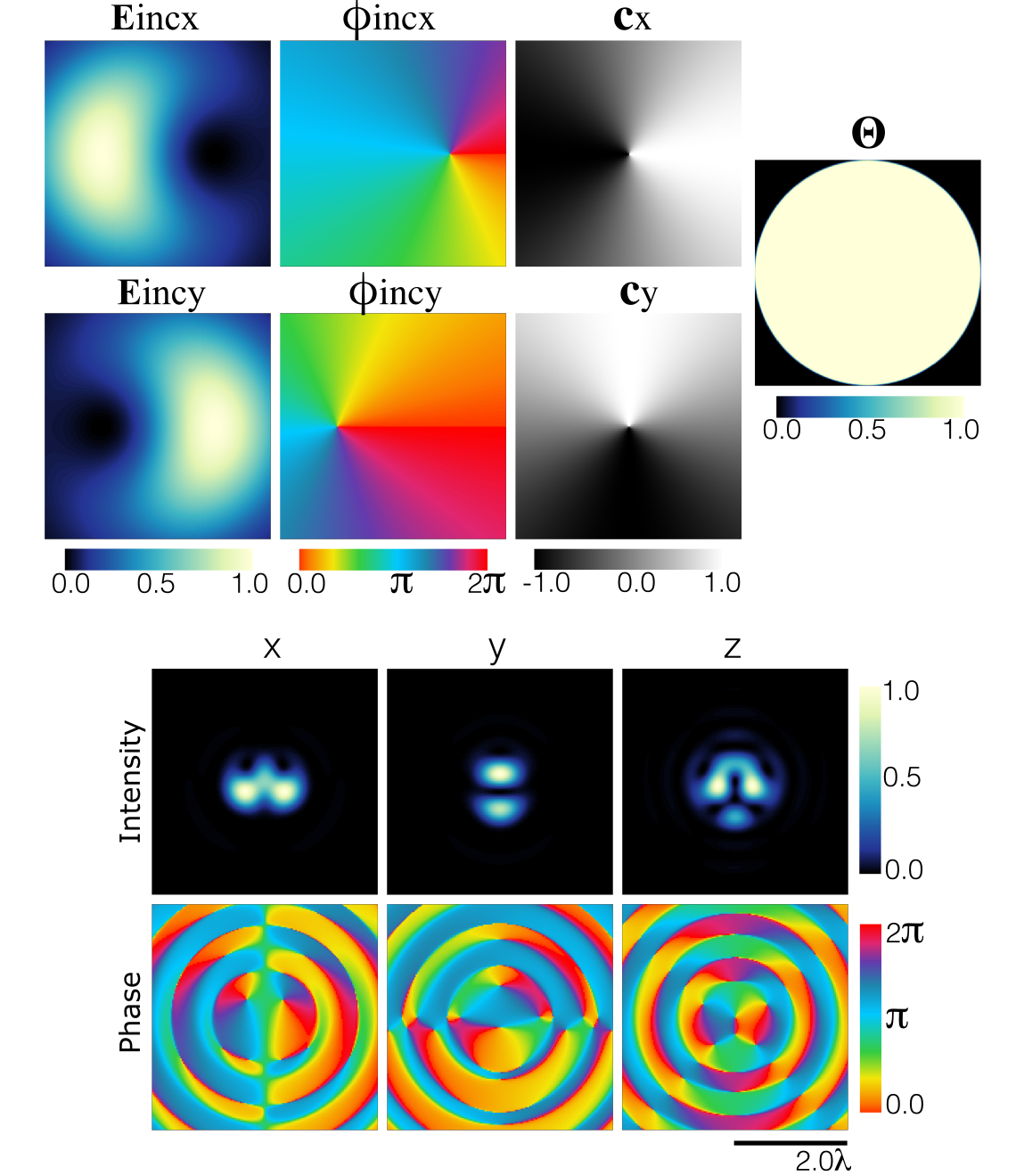}
    \caption{ Focusing of a beam with different amplitudes, phases, and polarization profiles in the transverse components. The first group of images shows the amplitude, phase, polarization maps and the circular aperture. The second group shows the focal fields. The parameters are $L=2^{12}$, $N=2^7$, $n_2=1.518$, NA$=1.3$, $f=2$ mm.   }
    \label{fig:arbitrary}
   \end{figure}

All the previous examples had considered $E_{0x}=E_{0y}$ and $\phi _{0x}=\phi _{0y}$. Here we consider a beam that has different amplitudes, phases, and polarization profiles in the transverse components. In Fig. \ref{fig:arbitrary} we show a beam with parameters in the angular coordinates:
\begin{equation}
\begin{split}   
E_{incx}(k_x,k_y) &=E_{0x} \sqrt{(k_x-k_{max}/2)^2+k_y^2} e^{\frac{-\left(k_x^2+k_y^2 \right)}{k_{max}^2}}\\
E_{incx}(k_x,k_y) &=E_{0y} \sqrt{(k_x+k_{max}/2)^2+k_y^2} e^{\frac{-\left(k_x^2+k_y^2 \right)}{k_{max}^2}}\\
\phi_{incx}&=\mathrm{tan}^{-1}[k_y/(k_x-k_{max}/2)]\\
\phi_{incy}&=\mathrm{tan}^{-1}[k_y/(k_x+k_{max}/2)]\\
c_{incx}&= k_x /\sqrt{k_x^2+k_y^2} \\
c_{incy}&= k_y /\sqrt{k_x^2+k_y^2} 
\end{split}
\end{equation}
Where $E_{0x}$ ($E_{0y}$) is such as the maximum value of $E_{incx}$ ($E_{incy}$) is one.  

\section{Planar Interface (propagating and evanescent)}
Many applications consider these beams propagating through a planar interface like optical micromanipulation where the beam propagates through a glass-liquid interface. 
Due to the refractive index mismatch across the planar boundary there is the possibility of total internal reflection and having evanescent waves.
In this section we reproduce the results contained in Novotny and Hetch book \cite{novotnybook}.
They consider that the field focuses at $z=0$ and the boundary ($xy$ plane) is at a height of $z_0$. The refractive index at $z<z_0$ is $n_1$ and $k_{z1}=\sqrt{k_1^2-(k_x^2+k_y^2)}$ ($k_1=n k_0$), above the boundary for $z>z_0$, the refractive index is $n_2$ and $k_{z2}=\sqrt{k_2^2-(k_x^2+k_y^2)}$ ($k_2=n_2 k_0$).

\begin{figure*} 
\begin{center} 
\includegraphics[width=1\textwidth]{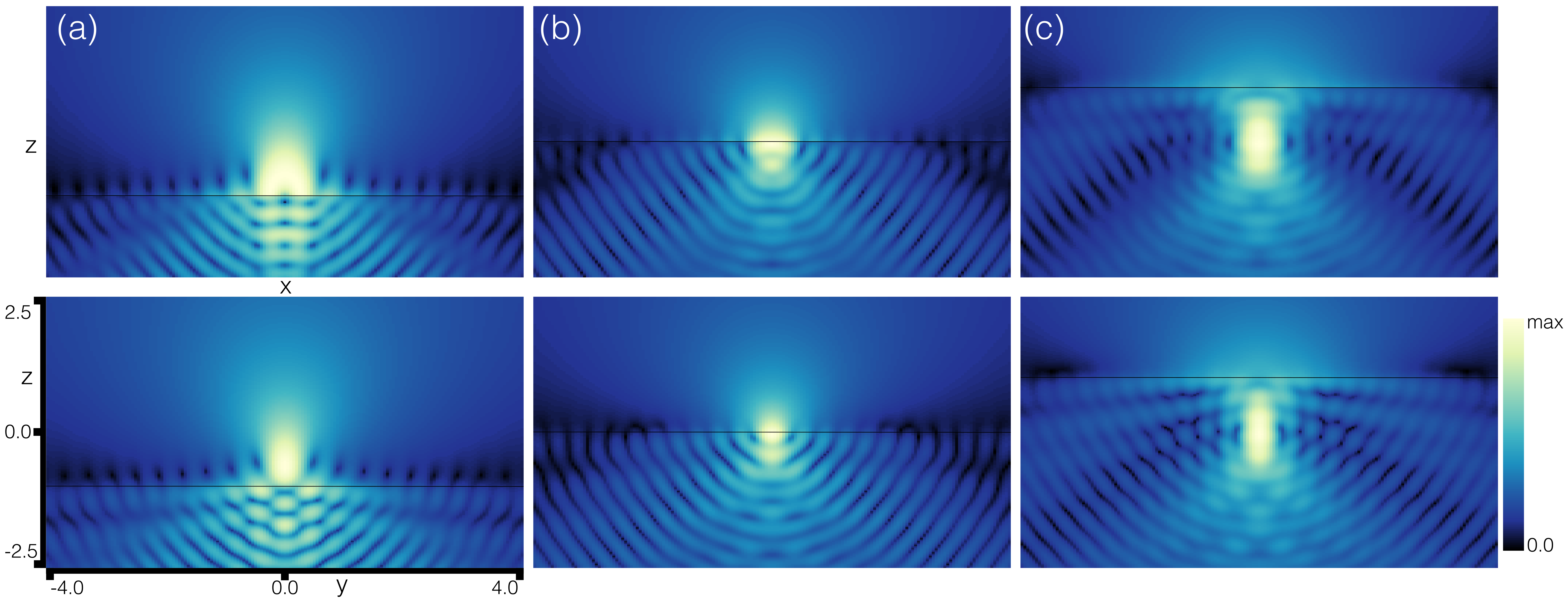}
\end{center}
\caption{Intensity $\textbf{E}\cdot\textbf{E}^*$ of a focused linearly polarized Gaussian beam with parameters $f_0=1$, NA$=1.4$, close to an interface glass-air. The spanned area $5\lambda \times 8 \lambda$ is the same for all the images, the $(0,0)$  is the focus of the lens. In the first row is shown the plane $(x,z)$ and in the second row is shown the plane $(y,z)$, where $z$. (a) The interface is located at $z_0=-\lambda$. (b) The interface is located at the focal plane $z_0=0$. (c) The interface is located at $z_0=\lambda$ and. We computed the field at 200 planes along the z direction}. In the three cases L=2048 and N=150. 
\label{fig:inter0}
\end{figure*}  

\subsection{Fresnel Coefficients}
The reflection and transmission Fresnel coefficients for a planar boundary (s and p components) are:

\begin{equation}
\begin{split}
    r^s (k_x, k_y)&=\frac{\mu_2 k_{z1}-\mu _1 k_{z2}}{\mu _2 k_{z1}+\mu _1 k_{z2}}
    \\
    r^p (k_x, k_y)&=\frac{\epsilon_2 k_{z1}-\epsilon _1 k_{z2}}{\epsilon _2 k_{z1}+\epsilon _1 k_{z2}}
    \\
    t^s(k_x, k_y)&= \frac{2\mu _2 k_{z1}}{\mu _2 k_{z1} +\mu _1 k_{z2}}
    \\
    t^p(k_x, k_y)&= \frac{2\epsilon _2 k_{z1}}{\epsilon _2 k_{z1} +\epsilon _1 k_{z2}} \sqrt{\frac{\mu _2 \epsilon _1}{ \mu _1 \epsilon _2}}
\end{split}
\end{equation}
Normally, for a dielectric we can set $\mu _1 =\mu _2 =1$. 

The field between the lens and the end of the boundary (same material) is a superposition of the focused and reflected fields $\textbf{E}=\textbf{E}_f(x,y,z)+\textbf{E}_r(x,y,z)$ for $z < z_0$:
\begin{align}
\textbf{E}_f(x,y,z)=-\frac{ife^{-ikf}}{2\pi}\iint\limits_{-\infty}^{\infty} \left[ c_x \textbf{E}_{\infty ,f}^{x}+c_y\textbf{E}_{\infty ,f}^{y} \right]\times  \nonumber\\
\frac{1}{k_z} e^{ik_x x+ik_y +ik_{z_1} z} dk_x dk_y
\end{align}
\begin{align}
\textbf{E}_r(x,y,z)=-\frac{ife^{-ikf}}{2\pi}\iint\limits_{-\infty}^{\infty} \left[ c_x \textbf{E}_{\infty ,r}^{x}+c_y\textbf{E}_{\infty ,r}^{y} \right] \times \nonumber \\
\frac{1}{k_z} e^{ik_x x+ik_y -ik_{z_1} z} dk_x dk_y
\end{align}
In the case of $z>z_0$ then $\textbf{E}=\textbf{E}_t$ where the transmitted field:
\begin{align}
\textbf{E}_t(x,y,z)=-\frac{ife^{-ikf}}{2\pi}\iint\limits_{-\infty}^{\infty} \left[ c_x \textbf{E}_{\infty ,t}^{x}+c_y\textbf{E}_{\infty ,t}^{y} \right]\times \nonumber \\
\frac{1}{k_z} e^{ik_x x+ik_y +ik_{z_2} z} dk_x dk_y
\end{align}

The reflected $\textbf{E}_{\infty ,r}$ and transmitted $\textbf{E}_{\infty ,t}$ in terms of the $xy$ components of $\textbf{E}_{inc}$: 
\begin{align}
\textbf{E}_{\infty , r}^x=\sqrt{\frac{k_{z_1}}{k_1}} E_{inc}^x(\frac{k_x}{k},\frac{k_y}{k}) \left( \frac{e^{2ik_{z_1}z_0}}{k_x^2+k_y^2} \right) \times \nonumber \\
\begin{pmatrix}
           r^s k_y^2- r^p k_x^2 k_{z_1}/k_1\\
            -r^s k_x k_y - r^p k_x k_y k_{z_1} /k_1\\
          - r^p (k_x^2+k_y^2)k_x/k_1)
         \end{pmatrix}  
\end{align}

For the other transverse polarization component (direction $\textbf{n}_y$): \\
\begin{align}
\textbf{E}_{\infty ,r}^{y}=\sqrt{\frac{k_{z_1}}{k_1}} E_{inc}^y(\frac{k_x}{k},\frac{k_y}{k})   \left( \frac{e^{2ik_{z_1}z_0}}{k_x^2+k_y^2} \right)\times \nonumber \\
\begin{pmatrix}
           -r^s k_x k_y- r^p k_x k_y k_z/k_1\\
            r^s k_x^2 - r^p k_y^2 k_{z1} /k_1\\
          - r^p (k_x^2+k_y^2)k_y/k_1)
         \end{pmatrix}  
\end{align}

The transmitted component:
\begin{align}
\textbf{E}_{\infty , t}^x=\sqrt{\frac{k_{z_1}}{k_1}} E_{inc}^x(\frac{k_x}{k},\frac{k_y}{k}) \left( \frac{e^{i(k_{z_1}-k_{z_2} ) z_0}}{k_x^2+k_y^2} \right)\times  \nonumber\\
\begin{pmatrix}
           t^s k_y^2+t^p k_x^2 k_{z_2}/k_2\\
            -t^s k_x k_y + t^p k_x k_y k_{z2} /k_2\\
           -t^p (k_x^2+k_y^2)k_x/k_2)
         \end{pmatrix}  \frac{k_{z_2}}{k_{z_1}}
\end{align}       

For the other transverse polarization component (direction $\textbf{n}_y$). \\
\begin{align}
\textbf{E}_{\infty , t}^{y}=\sqrt{\frac{k_{z_1}}{k_1}} E_{inc}^y(\frac{k_x}{k},\frac{k_y}{k})  \left( \frac{e^{i(k_{z_1}-k_{z_2} ) z_0}}{k_x^2+k_y^2} \right) \times \nonumber\\
\begin{pmatrix}
           -t^s k_x k_y+ t^p k_x k_y k_{z2}/k_2\\
            t^s k_x^2+ t^p k_y^2 k_{z2} /k_2\\
           -t^p (k_x^2+k_y^2)k_y/k_2)
         \end{pmatrix} \frac{k_{z_2}}{k_{z_1}} 
\end{align}

Figure 9 shows the case of the linearly polarized Gaussian with $w_0 =R$ ($f_0=1$) focused by a lens with $NA=1.4$ at $z=0$, the position of the interface is at $z_0=-\lambda\,,0\,,\lambda$ (columns 1, 2 and 3 respectively). The first row has cross sections of the plane $x,z$ with $y=0$ and the second row plane $y,z$ with $x=0$. The colormaps represent the fourth root of total intensity in order to show more clearly the details. The case with $z_0=0$ appears in \cite{novotnybook}.

\section{Conclusion}
We presented two simple programs to calculate tightly focused vectorial light fields: one for a propagating field and the other considers a planar interface where evanescent fields can emerge. 
These programs are based on previous results described in \cite{fastfocal, novotnybook}. The main contribution is the discussion about the correction that has to be made for a shifted Fourier transform, which is a subtle detail that is easy to miss.
Also, the use of Cartesian coordinates helps when making comparisons with experiments where the light is shaped by rectangular arrays which are described in \cite{novotnybook}.

\subsection*{Acknowledgements}
Work partially funded by DGAPA UNAM PAPIIT grants IN107719 and  IN107222; CTIC-LANMAC and CONACYT LN-299057.

%



\clearpage



\section*{Supplementary material}

# APPENDIX A: PROGRAM TO CALCULATE PROPAGATING BEAM

```matlab
%%%%%%%%%%%%%%%%%%%%%%%%%%%%%%%%%%%%%%%%%%%%%%%%%%%%%%%%%%%%%%%%%%%%
clear all; %close all; clc
%run time sim 10s with L=2^12, and N=2^7
%%%%%%%%%%%%%%%%%%%%%%%%%%%%%%%%%%%%%%%%%%%%%%%%%%%%%%%%%%%%%%%%%%%%
%%%%%%%%%%%%%%%%%initial parameters
L=2^12; % main domain size
N=2^7; % number of points for the aperture radius
z=0; %observation plane (0 at the focal plane)
m0=1/2; %coordinate shift
xmax=0.9; %half the size of the cropped output in wavelengths
%(final size of images is 2xmax)
n1=1.0;  % refractive index before the lens (air)
n2=1.518; % refractive index after the lens (immersion oil)
f=1800; % effective focal length of the lens in um (10^-6 m)
NA=1.4; %effective numerical aperture
lambda=1.064;  %wavelength in um
R=f*NA/n2;  %aperture radius in um;
dx_inf=R/N;  %spatial resolution at aperture um/px or um/point

%%%%%%%%%%%%%%%%%%%%%%%%%%%%%%%%%%%%%%%%%%%%%%%%%%%%%%%%%%%%%%%%%%%%
%%%%%%%%%% Spatial coordinate system before lens (x_inf, y_inf)
d_inf=linspace(-L/2+m0,L/2-m0,L)*dx_inf; %spatial axis shifted by m0.
[x_inf,y_inf]=meshgrid(d_inf,-d_inf); %mesh, y inverted in Matlab images
[theta_inf,rho_inf] = cart2pol(x_inf,y_inf); %auxiliary polar coordinates

%%%%%%%%%%%%%%%%%%%%%%%%%%%%%%%%%%%%%%%%%%%%%%%%%%%%%%%%%%%%%%%%%%%%
%%%%%%%%%%%%%% angular
k0=2*pi/lambda;
k1=n2*k0;
dk=k0*NA/N;    %dk=(1/2)*(k1/(f))*dx0;
kx=x_inf*k1/f;
ky=y_inf*k1/f;
dxf=2*pi/(L*dk); %equivalent to dxf=N*lambda/(L*NA);
%conversion factor at output plane in um/px
kM1=k1*ones(L,L);       %%%array with magnitude of k1
kz1=sqrt(kM1.^2-kx.^2-ky.^2);

%%%%%%%%%%%%%%%%%%%%%%%%%%%%%%%%%%%%%%%%%%%%%%%%%%%%%%%%%%%%%%%%%%%%
xmax2=round(xmax*lambda/dxf); %xmax in px at output plane

%%%%%%%%%%%%%%%%%%%%%%%%%%%%%%%%%%%%%%%%%%%%%%%%%%%%%%%%%%%%%%%%%%%%
%%%%%%%%%%%%%%%%%%%%%%%%correction for m0 shift
x=linspace(-L/2,L/2,L); %%%auxiliary coordinate for correction
y=x;  [X,Y]=meshgrid(x,y);
PhaseShift=m0*2*pi*X/L+m0*2*pi*Y/L; %correction phase
%Center of Fourier transform displaced (vert and hor) by m0.
%shift of \∆ kx=\∆ ky= m0*(2*pi/L).

%%%%%%%%%%%%%%%%%%%%%%%%%%%%%%%%%%%%%%%%%%%%%%%%%%%%%%%%%%%%%%%%%%%%
%%%%%%%%%%%%%%%%%%%%%%%%input field   initial field transverse
w0=f*NA/n2;  % The beam waist is equal to the aperture radius.
E_incx=exp(-(x_inf.^2+y_inf.^2)/(w0^2));  %normalized amplitude E_x0   A_0
phi_incx=double(zeros(L,L));   %%%constant
E_incy=E_incx;
phi_incy = phi_incx;

%%%%%%%%%%%%%%%%%%%%%%%%%%%%%%%%%%%%%%%%%%%%%%%%%%%%%%%%%%%%%%%%%%%%
%%%%%%%%%%%%%%%%round soft aperture (spatial coordinates)
Theta=0.5*(1+tanh((1.5/1)*(N-sqrt((x_inf/dx_inf).^2+(y_inf/dx_inf).^2))));
%angular space
%kmax=R*k1/f;
%Theta=0.5*(1+tanh((1.5/dk)*(kmax-sqrt((kx).^2+(ky).^2))));

%%%%%%%%%%%%%%%%%%%%%%%%%%%%%%%%%%%%%%%%%%%%%%%%%%%%%%%%%%%%%%%%%%%%
%%%polarization state
%cx=ones(L,L); cy=zeros(L,L);   % lin horizontal
```

```matlab
67  cx=zeros(L,L); cy=ones(L,L);   % lin vertical
68  %cx=1*ones(L,L); cy=1*ones(L,L);    % lin diag
69  %cx=ones(L,L); cy=1i*ones(L,L); %%right circ
70  %cx=1i*ones(L,L); cy=ones(L,L); %%left circ
71  %cx=kx./sqrt(kx.^2+ky.^2);  cy=ky./sqrt(kx.^2+ky.^2); %%radial
72  %cx=-ky./sqrt(kx.^2+ky.^2);  cy= kx./sqrt(kx.^2+ky.^2);%%azimuthal
73  %phi_aux=mod(atan2(ky,kx),2*pi); s=8; cx=cos((s/2)*phi_aux); cy=sin((s/2)*phi_aux); %flower
74
75
76
77  %%%%%%%%%%%%%%%%%%%%%%%%%%%%%%%%%%%%%%%%%%%%%%%%%%%%%%%%%%%%%%%%
78  %%%%%%%%E_inf   %%%%%%%%%%%%%%%%%%%%%%%%%%%%%%%%%%%%%%%%%%%%%%%%
79  %input x
80  %x component
81  E_infxx=sqrt(n1/n2)*sqrt(kz1./kM1).*Theta.*E_incx.*exp(1i*phi_incx).*...
82      (ky.^2+kx.^2.*kz1./kM1)./(kx.^2+ky.^2);
83  %y component
84  E_infxy=sqrt(n1/n2)*sqrt(kz1./kM1).*Theta.*E_incx.*exp(1i*phi_incx).*...
85      (-kx.*ky+kx.*ky.*kz1./kM1)./(kx.^2+ky.^2);
86  %z component
87  E_infxz=sqrt(n1/n2)*sqrt(kz1./kM1).*Theta.*E_incx.*exp(1i*phi_incx).*...
88      (-(kx.^2+ky.^2).*kx./kM1)./(kx.^2+ky.^2);
89
90  %y input
91  %x component
92  E_infyx=sqrt(n1/n2)*sqrt(kz1./kM1).*Theta.*E_incy.*exp(1i*phi_incy).*...
93      (-ky.*kx+kx.*ky.*kz1./kM1)./(kx.^2+ky.^2);
94  %y component
95  E_infyy=sqrt(n1/n2)*sqrt(kz1./kM1).*Theta.*E_incy.*exp(1i*phi_incy).*...
96      (kx.^2+ky.^2.*kz1./kM1)./(kx.^2+ky.^2);
97  %z component
98  E_infyz=sqrt(n1/n2)*sqrt(kz1./kM1).*Theta.*E_incy.*exp(1i*phi_incy).*...
99      (-(kx.^2+ky.^2).*ky./kM1)./(kx.^2+ky.^2);
100
101 %factors to assemble E_inf
102 CF2=-1i*f*exp(-1i*f*2*pi*n2/lambda)*exp(1i*kz1.*z)./(2*pi*kz1);
103
104 %%E_inf
105 Fieldx=CF2.*(cx.*E_infxx +cy.*E_infyx);
106 Fieldy=CF2.*(cx.*E_infxy +cy.*E_infyy);
107 Fieldz=CF2.*(cx.*E_infxz +cy.*E_infyz);
108
109 %%%%%%%%%%%%%%%%%%%%%%%%%%%%%%%%%%%%%%%%%%%%%%%%%%%%%%%%%%%%%%%%
110 %%%%%Integration
111 Ex0=ifftshift(ifft2(fftshift((Fieldx)))).*exp(1i*PhaseShift);
112 Ex=Ex0((L/2+1-xmax2):(L/2+1+xmax2),(L/2+1-xmax2):(L/2+1+xmax2)); %cropped
113 %x=y=0 at the center
114
115 Ey0=ifftshift(ifft2((fftshift(Fieldy)))).*exp(1i*PhaseShift);
116 Ey=Ey0((L/2+1-xmax2):(L/2+1+xmax2),(L/2+1-xmax2):(L/2+1+xmax2)); %cropped
117
118 Ez0=ifftshift(ifft2((fftshift(Fieldz)))).*exp(1i*PhaseShift);
119 Ez=Ez0((L/2+1-xmax2):(L/2+1+xmax2),(L/2+1-xmax2):(L/2+1+xmax2)); %cropped
120
121 %%%%%%%%%%%%%%%%%%%%%%%%%%%%%%%%%%%%%%%%%%%%%%%%%%%%%%%%%%%%%%%%
122 %%%%%%%%%%%%%%%%%%%%%%%%%%%%%%%%%%%%%%%%%%%%%%%%%%%%%%%%%%%%%%%%
123
124
125 %intensity ratios
126 c0y=max(max(abs(Ey).^2))/max(max(abs(Ex).^2));
127 c0z=max(max(abs(Ez).^2))/max(max(abs(Ex).^2));
128
129 %%%%%%%%%%%%%%%%%%%%%%%%%%%%%%%%%%%%%%%%%%%%%%%%%%%%%%%%%%%%%%%%
130 %plots
131
132 a='jet';
133 b='gray';
134
135 figure(1) %amplitude Ex
136 Exn=abs(Ex)./max(max(abs(Ex)));
```

# APPENDIX B: PROGRAM FOR INTERFACE

```matlab
%%%%%%%%%%%%%%%%%%%%%%%%%%%%%%%%%%%%%%%%%%%%%%%%%%%%%%%
clear all; %
%%%%%%%%%%%%%%%%%%%%%%%%%%%%%%%%%%%%%%%%%%%%%%%%%%%%%%%
%%%%%%%%%%%%%%%%initial parameters
L=2^10; % main domain size
N=2^6; % number of points for the aperture radius
z0=0; %interface position (focus at z=0)
z=linspace(-2,2,41); %axial domain
m0=1/2; %coordinate shift
xmax=4; %half the size of the cropped output in wavelengths
%(final size of images is 2xmax)
n1=1.518;  %refractive index for incident field (immersion oil)
n2=1.518; % refractive index for reflected field  (immersion oil)
n3=1; % refractive index for transmitted field (after interface)
e1=n2^2;
e2=n3^2;
m1=1; %permeability
m2=1; %permeability
f=1800; % effective focal length of the lens in um (10^-6 m)
NA=1.3; %effective numerical aperture
lambda=1.064;  %wavelength in um
l=1.064;  %wavelength   um
R=f*NA/n2;  %aperture radius in um;
dx_inf=R/N;  %spatial dimension at aperture um/px or um/point

%%%%%%%%%%%%%%%%%%%%%%%%%%%%%%%%%%%%%%%%%%%%%%%%%%%%%%%%%%%%%
%%%%%%%%% Spatial coordinate system before lens (x_inf, y_inf)
d_inf=linspace(-L/2+m0,L/2-m0,L)*dx_inf; %spatial axis shifted by m0.
[x_inf,y_inf]=meshgrid(d_inf,-d_inf); %mesh, y inverted in Matlab images
[theta_inf,rho_inf] = cart2pol(x_inf,y_inf); %auxiliary polar coordinates
%%%%%%%%%%%%%%%%%%%%%%%%%%%%%%%%%%%%%%%%%%%%%%%%%%%%%%%%%%%%%

%%%%%%%%%%%%%%%%%%%%%%%%%%%%%%%%%%%%%%%%%%%%%%%%%%%%%%%%%%%%%
%%%%%%%%%%%%% angular
k0=2*pi/lambda;
k1=n2*k0;
k2=n3*k0;
dk=k0*NA/N;     %dk=(1/2)*(k1/(f))*dx0;
kx=x_inf*k1/f;
ky=y_inf*k1/f;
dxf=2*pi/(L*dk); %equivalent to dxf=N*lambda/(L*NA);
%conversion factor at output plane in um/px
kM1=k1*ones(L,L);       %%%array with magnitude of k1
kz1=sqrt(kM1.^2-kx.^2-ky.^2);
kM2=k2*ones(L,L);
kz2=sqrt(kM2.^2-kx.^2-ky.^2);

%%%%%%%%%%%%%%%%%%%%%%%%%%%%%%%%%%%%%%%%%%%%%%%%%%%%%%%%%%%%%
xmax2=round(xmax*lambda/dxf); %xmax in px at output plane

%%%%%%%%%%%%%%%%%%%%%%%%%%%%%%%%%%%%%%%%%%%%%%%%%%%%%%%%%%%%%
%%%%%%%%%%%%%%%%%%%%%%correction for m0 shift
x=linspace(-L/2,L/2,L); %%%auxiliary coordinate for correction
y=x; [X,Y]=meshgrid(x,y);
PhaseShift=m0*2*pi*X/L+m0*2*pi*Y/L; %correction phase
%Center of Fourier transform displaced (vert and hor) by m0.
%shift of \∆ kx=\∆ ky= m0*(2*pi/L).

%%%%%%%%%%%%%%%%%%%%%%%%%%%%%%%%%%%%%%%%%%%%%%%%%%%%%%%%%%%%%
%%%%%%%%%%%%%%%%%%%%%%%input field    initial field transverse
w0=f*NA/n2;  % The beam waist is equal to the aperture radius.
E_incx=exp(-(x_inf.^2+y_inf.^2)/(w0^2));  %normalized amplitude E_x0   A_0
phi_incx=double(zeros(L,L));  %%%phase phi_x0 fase init;   phi_0
E_incy=E_incx;
phi_incy = phi_incx;

```

```matlab
%%%%%%%%%%%%%%%%%%%%%%%%%%%%%%%%%%%%%%%%%%%%%%%%%%%%%%%%%%%%%%%%%%%%
%%%%%%%%%%%%%round soft aperture (spatial coordinates)
Theta=0.5*(1+tanh((1.5/1)*(N-sqrt((x_inf/dx_inf).^2+(y_inf/dx_inf).^2))));
%angular space
%kmax=R*k1/f;
%Theta=0.5*(1+tanh((1.5/dk)*(kmax-sqrt((kx).^2+(ky).^2))));


%%%%%%%%%%%%%%%%%%%%%%%%%%%
%%%polarization
cx=ones(L,L); cy=zeros(L,L);   % lin horizontal
%cx=zeros(L,L); cy=ones(L,L);   % lin vertical
%cx=1*ones(L,L); cy=1*ones(L,L);    % lin diag
%cx=ones(L,L); cy=1i*ones(L,L); %%right circ
%cx=1i*ones(L,L); cy=ones(L,L); %%left circ
%cx=(kx)./sqrt(kx.^2+ky.^2);  cy=(ky)./sqrt(kx.^2+ky.^2); %%radial   cx=cos(phi) cy=sin(phi)
%cx=-ky./sqrt(kx.^2+ky.^2);  cy= kx./sqrt(kx.^2+ky.^2);%%azimuthal   cx=cos(phi) cy=sin(phi)
%phik=atan2(ky,kx); s12=8; cx=cos((s12/2)*phik); cy=sin((s12/2)*phik);%flower
%phik=atan2(ky,kx); s12=-8; cx=cos((s12/2)*phik); cy=sin((s12/2)*phik);% spider

%%%%%%%%%%%%%%%%%%%%%%%%%%%%%%%%%%%%%
%%Fresnel coeficients
rs=(m2*kz1-m1*kz2)./(m2*kz1+m1*kz2);
rp=(e2*kz1-e1*kz2)./(e2*kz1+e1*kz2);
ts=(2*m2*kz1)./(m2*kz1+m1*kz2);
tp=sqrt((m2*e1)/(m1*e2))*(2*e2*kz1)./(e2*kz1+e1*kz2);

%%%%%%%%%%%%%%%%%%%%%%%%%%%%%%%%%%%%%%%%%%%%%%%%%%%
%%%%%Factors for E_inf
CF1=-(1./kz1).*sqrt(kz1./kM1).*(1./(kx.^2+ky.^2))*1i*f*exp(-1i*f*2*pi*n2/lambda); %Factors for ...
    E_inf_r and f
CF2=-(1./kz2).*sqrt(kz1./kM1).*(1./(kx.^2+ky.^2))*1i*f*exp(-1i*f*2*pi*n2/lambda).*(kz2./kz1); ...
    %Factors for E_inf _t

%%%%%%%%%%%%%%%%%%%%%%%%%%%%%%%%%%%%%%%%%%%%%%%%%%%%%%%%%%%%%%%%%%%5
%%%%%%%%%%%%Focused propagating before interface
%for E_inf
% x component
MTx1=(ky.^2 +kx.^2.*kz1./kM1).*CF1; %lens transmission factor, incident x
MTx2=(-ky.*kx +(ky.*kx.*kz1)./kM1).*CF1; %lens transmission factor incident y
%y component
MTy1=(-kx.*ky+(kx.*ky.*kz1)./kM1).*CF1;%lens transmission factor, incident x
MTy2=(kx.^2 +ky.^2.*kz1./kM1).*CF1;%lens transmission factor, incident y
%z component
MTz1=(-(kx.^2+ky.^2).*kx./kM1).*CF1;%lens transmission factor, incident x
MTz2=(-(kx.^2+ky.^2).*ky./kM1).*CF1; %lens transmision factor, incident y

%%%%%%%%%%%%%%%%%%%%%%%%%%%%%%%%%%%%%%%%%%%%%%%%%%%%%%%%%%%%
%%%%%%%%%%%% Reflected E_inf_r
%x component
MTx1r=(rs.*ky.^2 -rp.*kx.^2.*kz1./kM1).*CF1; % incident x
MTx2r=(-rs.*ky.*kx  -rp.*(ky.*kx.*kz1)./kM1).*CF1;% incident y
%y component
MTy1r=(-rs.*kx.*ky-rp.*(kx.*ky.*kz1)./kM1).*CF1;% incident x
MTy2r=(rs.*kx.^2 -rp.*ky.^2.*kz1./kM1).*CF1;% incident y
%z component
MTz1r=(-rp.*(kx.^2+ky.^2).*kx./kM1).*CF1;%incident x
MTz2r=(-rp.*(kx.^2+ky.^2).*ky./kM1).*CF1; % incident y

%%%%%%%%%%%%%%%%%%%%%%%%%%%%%%%%%%%%%%%%%%%%%%%%%%%%%%%%%%%%
%%%%%%%%%%%%Transmited E_inf_t
%x component
MTx1t=(ts.*ky.^2+tp.*kx.^2.*kz2./kM2).*CF2; % x input
MTx2t=(-ts.*ky.*kx+tp.*(ky.*kx.*kz2)./kM2).*CF2;% y input
%y component
MTy1t=(-ts.*kx.*ky+tp.*(kx.*ky.*kz2).*(1./kM2)).*CF2;% x input
MTy2t=(ts.*kx.^2+tp.*ky.^2.*kz2./kM2).*CF2;% y input
%z component
MTz1t=(-tp.*(kx.^2+ky.^2).*kx./kM2).*CF2;%x input
```

```matlab
MTz2t=(-tp.*(kx.^2+ky.^2).*ky./kM2).*CF2; % y input

%%%%%%%%%%%%%%%%%%%%%%%%%%%%%%%%%%%%%%%%%%%%%%%%%%%%%%%%%%%%%%%
%%%%%%%%%%%%%start

for j=1:length(z);
Hzf=exp(1i*kz1.*z(j)).*Theta;   %propagating
Hzr=exp(-1i*kz1.*z(j)+1i*2*kz1*z0).*Theta; %reflected
Hzt=exp(1i*kz2.*z(j)+1i*(kz1-kz2)*z0).*Theta; %transmitted



if z(j)≤z0;   % propagating and reflected up to the boundary at z<z0
%%%%%%%%%% focal
% x component E_inf
Fieldxxf=(cx.*MTx1.*E_incx.*exp(1i*phi_incx)+cy.*MTx2.*E_incy.*exp(1i*phi_incy)).*Hzf;
%integration x component
Ex_inff=ifftshift(ifft2(fftshift((Fieldxxf)))).*exp(1i*PhaseShift);
Exf=Ex_inff((L/2+1-xmax2):(L/2+1+xmax2),(L/2+1-xmax2):(L/2+1+xmax2)); %cropped

%y component E_inf
Fieldxyf=(cx.*MTy1.*E_incx.*exp(1i*phi_incx)+cy.*MTy2.*E_incy.*exp(1i*phi_incy)).*Hzf;
%integration y component
Ey_inff=ifftshift(ifft2((fftshift(Fieldxyf)))).*exp(1i*PhaseShift);
Eyf=Ey_inff((L/2+1-xmax2):(L/2+1+xmax2),(L/2+1-xmax2):(L/2+1+xmax2));

%z component E_inf
Fieldxzf=(cx.*MTz1.*E_incx.*exp(1i*phi_incx)+cy.*MTz2.*E_incy.*exp(1i*phi_incy)).*Hzf;
%integration z component
Ez0f=ifftshift(ifft2((fftshift(Fieldxzf)))).*exp(1i*PhaseShift);
Ezf=Ez0f((L/2+1-xmax2):(L/2+1+xmax2),(L/2+1-xmax2):(L/2+1+xmax2));
%%%%%%%%%%

%%%%%%%%%%%%%%%%%%%%%%%%%%%%%%%%%%%%%%%%%%%%%%%%%%%%%%%%
%%%%%%%%%% reflected
%x component E_inf_r
Fieldxxr=(cx.*MTx1r.*E_incx.*exp(1i*phi_incx)+cy.*MTx2r.*E_incy.*exp(1i*phi_incy)).*Hzr;
%integration
Ex_infr=ifftshift(ifft2(fftshift((Fieldxxr)))).*exp(1i*PhaseShift);
Exr=Ex_infr((L/2+1-xmax2):(L/2+1+xmax2),(L/2+1-xmax2):(L/2+1+xmax2));

%y component E_inf_r
Fieldxyr=(cx.*MTy1r.*E_incx.*exp(1i*phi_incx)+cy.*MTy2r.*E_incy.*exp(1i*phi_incy)).*Hzr;
%integration
Ey_infr=ifftshift(ifft2((fftshift(Fieldxyr)))).*exp(1i*PhaseShift);
Eyr=Ey_infr((L/2+1-xmax2):(L/2+1+xmax2),(L/2+1-xmax2):(L/2+1+xmax2));

%z component E_inf_r
Fieldxzr=(cx.*MTz1r.*E_incx.*exp(1i*phi_incx)+cy.*MTz2r.*E_incy.*exp(1i*phi_incy)).*Hzr;
%integration
Ez0r=ifftshift(ifft2((fftshift(Fieldxzr)))).*exp(1i*PhaseShift);
Ezr=Ez0r((L/2+1-xmax2):(L/2+1+xmax2),(L/2+1-xmax2):(L/2+1+xmax2));

%%%%%%%%%%%%%%%%%%%%%%%%%%%%%%%%%%%%%%%%%%%%%%%%%%%%%
Ex(:,:,j)=Exf+Exr;   % total field at the material n_2 incident+reflected
Ey(:,:,j)=Eyf+Eyr;
Ez(:,:,j)=Ezf+Ezr;

%%%%%%%%%%
%end

%%%%%%%%%%%%%%%%%%%%%%%%%%%%%%%%%%%%%%%%%%%%%%%%%%%%%%%%
%%%%transmitted
else
% x component E_inf_t
Fieldxxt=(cx.*MTx1t.*E_incx.*exp(1i*phi_incx)+cy.*MTx2t.*E_incy.*exp(1i*phi_incy)).*Hzt;
%integration
Ex_inft=ifftshift(ifft2(fftshift((Fieldxxt)))).*exp(1i*PhaseShift);
Ext=Ex_inft((L/2+1-xmax2):(L/2+1+xmax2),(L/2+1-xmax2):(L/2+1+xmax2)); %cropped
```

```matlab
%y component E_inf_t
Fieldxyt=(cx.*MTy1t.*E_incx.*exp(1i*phi_incx)+cy.*MTy2t.*E_incy.*exp(1i*phi_incy)).*Hzt;
%integration
Ey_inft=ifftshift(ifft2((fftshift(Fieldxyt)))).*exp(1i*PhaseShift);
Eyt=Ey_inft((L/2+1-xmax2):(L/2+1+xmax2),(L/2+1-xmax2):(L/2+1+xmax2)); %cropped

%z component E_inf_t
Fieldxzt=(cx.*MTz1t.*E_incx.*exp(1i*phi_incx)+cy.*MTz2t.*E_incy.*exp(1i*phi_incy)).*Hzt;
%integration
Ez0t=ifftshift(ifft2((fftshift(Fieldxzt)))).*exp(1i*PhaseShift);
Ezt=Ez0t((L/2+1-xmax2):(L/2+1+xmax2),(L/2+1-xmax2):(L/2+1+xmax2)); %cropped

Ex(:,:,j)=Ext;   %field beyond the boundary, transmitted
Ey(:,:,j)=Eyt;
Ez(:,:,j)=Ezt;

end

end


%%%plot yz cross section through the center at xmax2+1
figure(1)
EEx=permute(Ex(xmax2+1,:,:),[3 2 1]);
EEy=permute(Ey(xmax2+1,:,:),[3 2 1]);
EEz=permute(Ez(xmax2+1,:,:),[3 2 1]);
I=abs(EEx).^2+abs(EEy).^2+abs(EEz).^2;
Imax=max(max(I));
In=I/Imax;
imagesc((In).^(1/4)); colormap(jet); colorbar
```




\begin{thebibliography}{0}%
\makeatletter
\providecommand \@ifxundefined [1]{%
 \@ifx{#1\undefined}
}%
\providecommand \@ifnum [1]{%
 \ifnum #1\expandafter \@firstoftwo
 \else \expandafter \@secondoftwo
 \fi
}%
\providecommand \@ifx [1]{%
 \ifx #1\expandafter \@firstoftwo
 \else \expandafter \@secondoftwo
 \fi
}%
\providecommand \natexlab [1]{#1}%
\providecommand \enquote  [1]{``#1''}%
\providecommand \bibnamefont  [1]{#1}%
\providecommand \bibfnamefont [1]{#1}%
\providecommand \citenamefont [1]{#1}%
\providecommand \href@noop [0]{\@secondoftwo}%
\providecommand \href [0]{\begingroup \@sanitize@url \@href}%
\providecommand \@href[1]{\@@startlink{#1}\@@href}%
\providecommand \@@href[1]{\endgroup#1\@@endlink}%
\providecommand \@sanitize@url [0]{\catcode `\\12\catcode `\$12\catcode
  `\&12\catcode `\#12\catcode `\^12\catcode `\_12\catcode `\%12\relax}%
\providecommand \@@startlink[1]{}%
\providecommand \@@endlink[0]{}%
\providecommand \url  [0]{\begingroup\@sanitize@url \@url }%
\providecommand \@url [1]{\endgroup\@href {#1}{\urlprefix }}%
\providecommand \urlprefix  [0]{URL }%
\providecommand \Eprint [0]{\href }%
\providecommand \doibase [0]{http://dx.doi.org/}%
\providecommand \selectlanguage [0]{\@gobble}%
\providecommand \bibinfo  [0]{\@secondoftwo}%
\providecommand \bibfield  [0]{\@secondoftwo}%
\providecommand \translation [1]{[#1]}%
\providecommand \BibitemOpen [0]{}%
\providecommand \bibitemStop [0]{}%
\providecommand \bibitemNoStop [0]{.\EOS\space}%
\providecommand \EOS [0]{\spacefactor3000\relax}%
\providecommand \BibitemShut  [1]{\csname bibitem#1\endcsname}%
\let\auto@bib@innerbib\@empty
\end{thebibliography}

\begin{thebibliography}{}
\bibitem{rw} B.Richards and E. Wolf, Electromagnetic diffraction in optical systems, II. Structure of the image field in an aplanatic system. Proc. R. Soc. Lond.- A. \textbf{253,} 358–379 (1959).
\bibitem{fastfocal} M. Leutenegger, R. Rao, R. A. Leitgeb, and T. Lasser. Fast focus field calculations. Opt. Express \textbf{14,} 11277-11291 (2006).
\bibitem{infocus} Q. Li, M. Chambonneau, M. Blothe, H. Gross, and S. Nolte, Flexible, fast, and benchmarked vectorial model for focused laser beams. Appl. Opt. 60, 3954-3963 (2021). 
\bibitem{pyfocus} F.Caprile, L. A. Masullo, F. D. Stefani, PyFocus – A Python package for vectorial calculations of focused optical fields under realistic conditions. Application to toroidal foci. Comput. Phys. Commun. \textbf{275}, 108315 (2022). 
\bibitem{fastfocalcartesian} Boruah, B. R. and Neil, M. A. A. Focal field computation of an arbitrarily polarized beam using fast Fourier transforms. Opt. Commun. 282, 4660–4667 (2009).
\bibitem{novotnybook} L. Novotny, and  Hecht. \textit{Principles of Nano-Optics.}  (Cambridge Univ. Press, 2012).
\bibitem{denz} E. Otte, K. Tekce, and C. Denz. Tailored intensity landscapes by tight focusing of singular vector beams, Opt. Express \textsf{25,} 20194-20201 (2017).
\end{thebibliography}
\end{document}